\documentclass{aa}
\usepackage{graphicx}
\usepackage{psfig}
\usepackage{times}
\def\lfir{$L_{\rm FIR}$}
\def\mabs{$M_{\rm B}$}

\def\fhi{$F_{\rm HI}$}
\def\lvm{$\log(V_{\rm m})$}
\def\vm{$V_{\rm m}$}

\def\micron{$\mu$m}
\def\jykms{Jy km s$^{-1}$}
\def\kms{km s$^{-1}$}
\def\kmsmpc{km s$^{-1}$ Mpc$^{-1}$}

\def\lsun{L$_{\odot}$}
\def\msun{M$_{\odot}$}

\begin{document}
\title{Starbursts in barred spiral galaxies}

\subtitle{VI. HI observations and the $K$-band Tully-Fisher relation
\thanks{Based on observations obtained at the large radiotelescope of 
Observatoire de Nan\c cay, operated by Observatoire de Paris}}

\author{E. Davoust\inst{}
\and
T. Contini\inst{}
}

\offprints{Davoust, \email{davoust@obs-mip.fr}}

\institute{
     UMR 5572, Observatoire Midi--Pyr\'en\'ees, 14 avenue E. Belin, F-31400 Toulouse, France
     }

\date{Received ; accepted }

\abstract{
This paper is primarily a study of the effect of a bar on the neutral hydrogen 
(HI) content of starburst and Seyfert galaxies. We also make comparisons 
with a sample of ``normal'' galaxies and investigate how well starburst
and Seyfert galaxies follow the fundamental scaling Tully-Fisher (TF) relation
defined for normal galaxies. 111 Markarian (Mrk) IRAS galaxies were observed
at the Nan\c cay radiotelescope, and HI data were obtained
for 80 galaxies,  of which 64 are new
detections. We determined the (20 and 50\%) linewidths, the 
maximum velocity of rotation and total HI flux for each galaxy. 
These measurements are complemented by data from the
literature to form a sample of Mrk IRAS (74\% starburst, 23\% 
Seyfert and 3\% unknown) galaxies containing 105 unbarred and 113 barred ones.  
Barred galaxies have lower total and bias-corrected HI masses
than unbarred galaxies, and this is true for both Mrk IRAS and
normal galaxies. This robust result suggests that bars funnel
the HI gas toward the center of the galaxy where it
becomes molecular before forming new stars.
The Mrk IRAS galaxies have higher bias-corrected 
HI masses than normal galaxies.
They also show significant departures 
from the TF relation, both in the $B$ and $K$ bands.  The most 
deviant points from the TF relation tend to have a strong far-infrared 
luminosity and a low oxygen abundance.  These results suggest that a 
fraction of our Mrk IRAS galaxies are still in the process of formation, 
and that their neutral HI gas, partly of external origin,
has not yet reached a stationary state. 
}

\maketitle
%

\section{Introduction}

The neutral hydrogen (HI) content of galaxies is a key 
parameter for the study of their evolution, because the HI gas is both the
reservoir for future star formation and an excellent tracer of the 
large-scale galaxy dynamics.

It has often been suggested that the nuclear activity in galaxies is
fuelled by accretion of gas toward the central regions, and 
that the mechanism for this radial infall is either 
a gravitational interaction or the presence of a bar (e.g., Noguchi
1988; Shlosman et al. 1989; Mihos \& Hernquist 1994; Friedli \& Benz 
1993, 1995).  The effect of a bar on nuclear activity
has been the main motivation behind the present series of papers,
and the first objective of this paper is to determine whether there is 
a difference in HI content between barred and unbarred galaxies. 

The presence of nuclear activity should also be related to
the overall gaseous content of the
host galaxies, since the availability of interstellar matter is
expected to affect 
the fueling rate of the nuclear regions.  So far,
only loose correlations have been found between the HI gas
content and the level of star formation in various samples of 
galaxies (e.g., Jackson et al. 1987; Mirabel \& Sanders 1988; 
Martin et al. 1991; Eskridge \& Pogge 1991; Andreani, Casoli \& G\'erin 1995; 
Contini 1996). A much tighter correlation has been found 
between the star formation rate (SFR) per unit area and the HI surface 
density in galaxies, extending over several orders of magnitudes in 
SFR and gas density (Kennicutt 1998).  
Our second objective is to investigate whether starburst 
and Seyfert galaxies have an abnormal neutral hydrogen content with 
respect to normal ones.

Another important use of HI data is for exploring the Tully-Fisher 
relation (Tully \& Fisher 1977; hereafter TF) for active galaxies. 
The TF relation is an empirical correlation which predicts that 
the absolute magnitude of a disk galaxy is proportional to its maximum 
rotational velocity \vm. In spite of the frequent use of the TF
relation as a distance indicator, the physical origin of this
relationship is still relatively poorly understood, and it remains
unclear whether all rotationally supported disk galaxies obey a single 
TF relation. As the fundamental scaling relation for spiral galaxies, 
the TF relation provides constraints on
galaxy formation, because it is deeply connected to the processes by which
disk galaxies form (e.g., Burstein \& Sarazin 1983; Cole et al. 1994; 
Eisenstein \& Loeb 1996; Steinmetz \& Navarro 1999; Koda, Sofue \& 
Wada 2000; Navarro \& Steinmetz 2000; van den Bosch 2000; Mo \& Mao
2000).

Until now, the primary
goal of detailed TF studies has been to establish the tightest
relation possible for use as a distance indicator. Thus, most local TF 
studies are limited to normal isolated galaxies, excluding
galaxies with nuclear activity (starburst or AGN) and with signs of 
interactions and/or tidal distorsion (e.g., Rubin et al. 1985; Pierce
\& Tully 1992; Courteau 1997; Tully \& Pierce 2000). 
The TF relation has not yet been extensively used to probe galaxy 
evolution. 
Coziol et al. (2000) have recently shown that a significant fraction 
of the Markarian (Mrk) starburst galaxies strongly deviate from the $B$-band 
TF relation defined for normal galaxies, suggesting that the disks 
of these galaxies are not in a state of dynamical equilibrium. 

The third goal of this paper is to investigate whether starburst
and Seyfert galaxies deviate systematically from the TF relation using $K$-band 
photometry, which is a better probe of galaxy masses than the $B$ band.
Our study complements searches for TF deviations in
low-mass galaxies (Courteau \& Rix 1999; O'Neil, Bothun \& Schombert 
2000; Hunter, Hunsberger \& Roye 2000; McGaugh et al. 2000), 
extreme late-type galaxies (Matthews, van Driel \& Gallagher 1998), 
asymmetric galaxies (Zaritsky \& Rix 1997) and galaxies in close pairs 
(Barton et al. 2001).

To address the goals outlined above, we have culled a sample of 
galaxies from the catalogue of Markarian galaxies,
retaining only those galaxies which had been detected 
by IRAS (in other words, those with a measured far-infrared flux). 

The outline of the paper is as follows. The samples of barred and unbarred
Mrk FIR-bright galaxies are described in Sect.~\ref{samples}. 
The HI observations and data reduction are presented in
Sect.~\ref{obs} and the results in Sect.~\ref{results}. 
A comparative analysis of the HI content in Mrk IRAS vs. normal, and of
barred vs. unbarred galaxies is given in Sect.~\ref{hicontent}. 
The behavior of Mrk IRAS galaxies in the TF plane is investigated in 
Sect.~\ref{tfrelation}. Our principal conclusions are summarized and 
interpreted in terms of evolutionary stage of starburst galaxies in 
Sect.~\ref{conclu}. Throughout this paper, all 
calculations assume an $\Omega = 1$ and $H_0=75$ \kmsmpc\ cosmology.


\begin{figure*}[t]
\centering
\vspace{-0.5cm}
\hspace{-3.0cm}
\includegraphics[width=21cm]{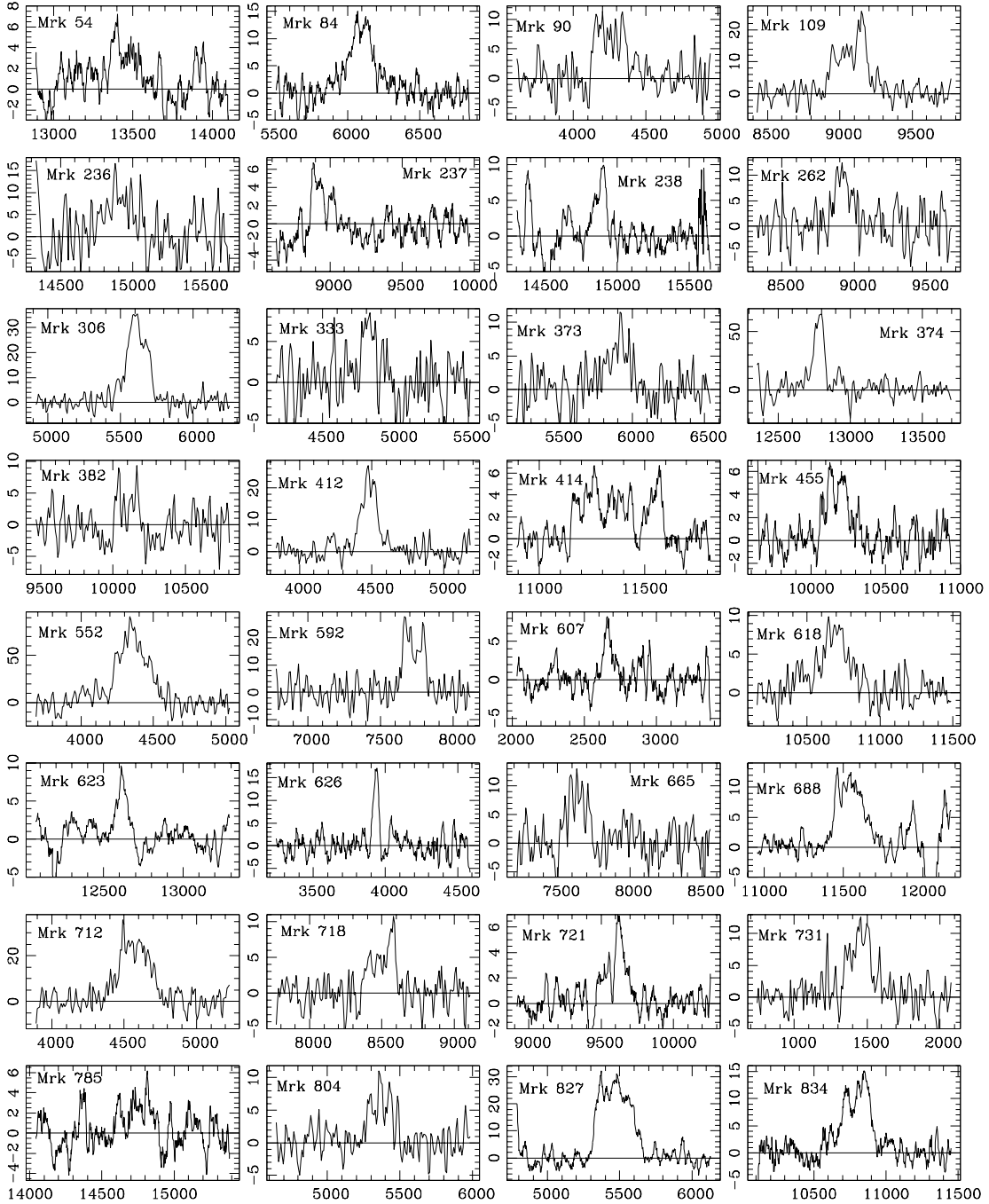}
   \vspace{-2cm}
      \caption{HI profiles for the Mrk IRAS
        galaxies. The $x$-axis is the heliocentric velocity in \kms;
        the $y$-axis is uncorrected flux density in mJy.}
         \label{hiprof1}
   \end{figure*}

   \begin{figure*}[t]
   \centering
   \vspace{-0.5cm}
    \hspace{-3.0cm}
   \includegraphics[width=21cm]{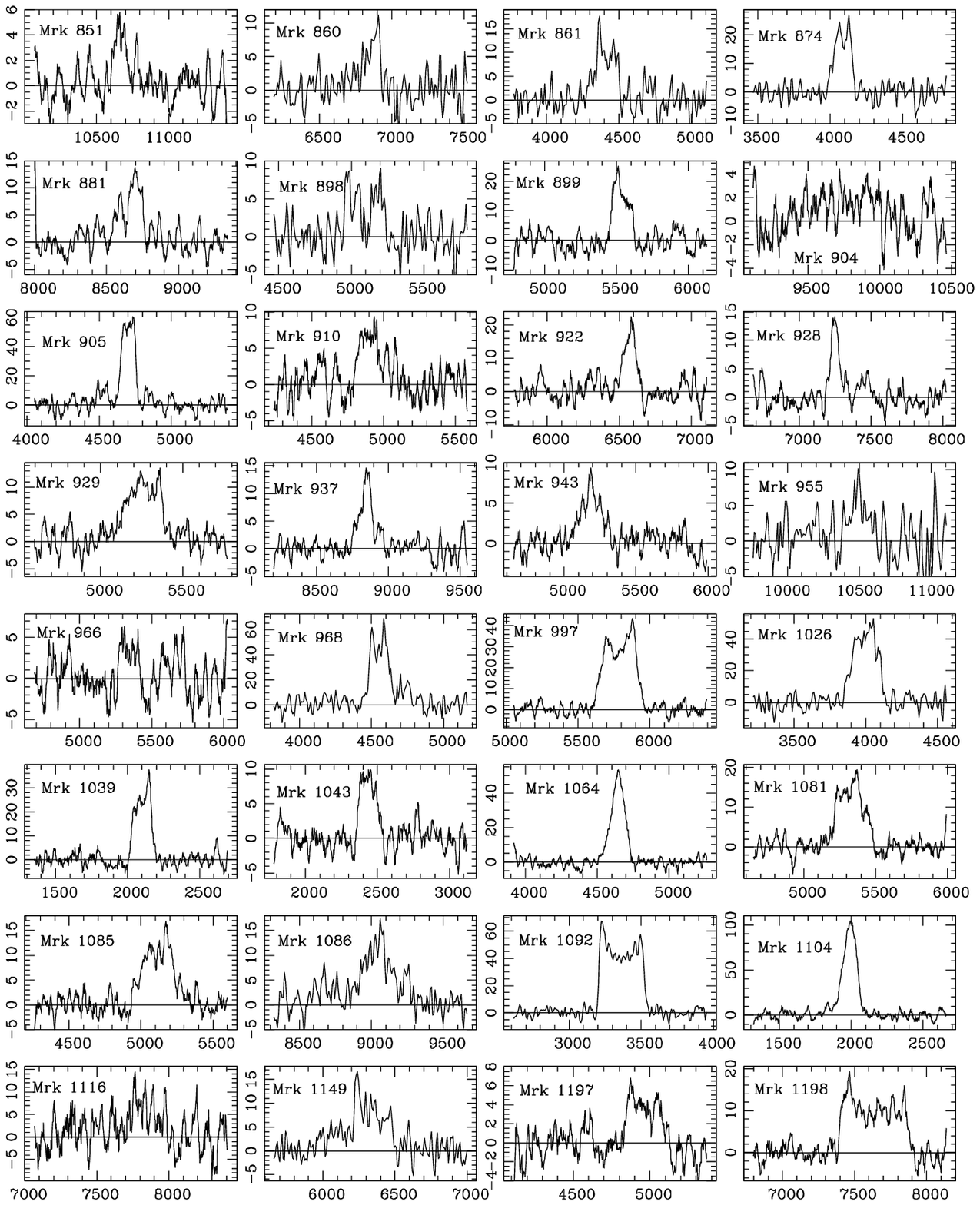}
   \vspace{-2cm}
      \caption{HI profiles for the Mrk IRAS
        galaxies ({\it continued}).}
         \label{hiprof2}
   \end{figure*}

   \begin{figure*}[t]
   \centering
   \vspace{-0.5cm}
    \hspace{-3.0cm}
   \includegraphics[width=21cm]{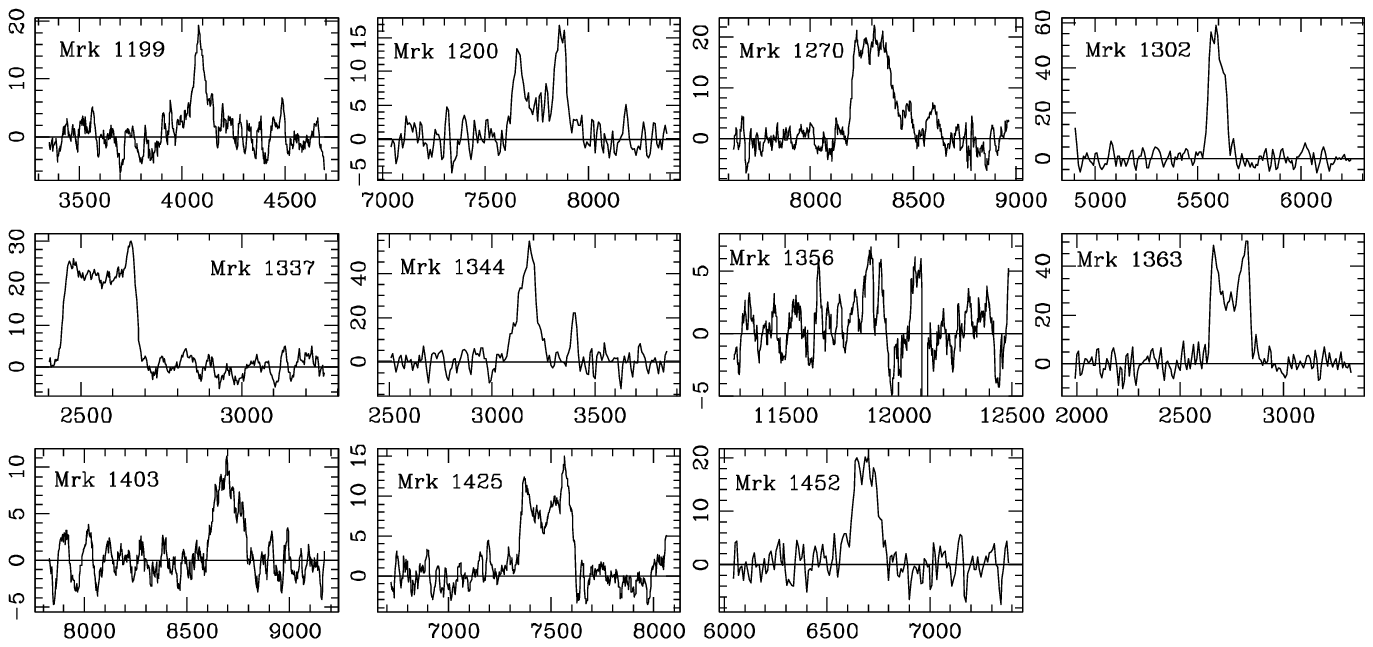}
   \vspace{-13cm}
      \caption{HI profiles for the Mrk IRAS
        galaxies ({\it continued}).}
         \label{hiprof3}
   \end{figure*}

\section{The two subsamples}
\label{samples}

The sample of Mrk IRAS galaxies was broken into two
subsamples, barred and unbarred galaxies, according to the morphological type
listed in LEDA\footnote{LEDA (http://leda.univ-lyon1.fr)}.  

The subsample of 144 barred galaxies contains $\sim$ 80\% of starbursts 
and 20\% of Seyfert galaxies.
It has been the subject of detailed multi-wavelength studies
(Contini 1996; Contini et al. 1995, 1997a,b, 1998; Chapelon et al.
1999; Coziol et al. 1997, 1999) and is described in more details 
by Contini et al. (1998).

The subsample of unbarred galaxies contains 110 Mrk IRAS galaxies with 
nearly the same relative proportion of starburst and Seyfert galaxies
as in the barred sample.  It is smaller than the barred 
subsample, but inspection of high-resolution images of the 
latter subsample (Contini 1996) revealed that 15 galaxies were in fact of
unbarred type, so that the two groups are in fact of comparable sizes.

\section{Observations and data reduction}
\label{obs}

When a literature search for HI observations of our sample turned
out a rather small number of galaxies, we decided to obtain HI
observations of the Mrk IRAS galaxies that had not yet been observed.

The observations were obtained at the large 
decimetric radiotelescope of Nan\c cay
Observatory\footnote{http://www.obs-nancay.fr/html\_an/a\_tecrt.htm}. 
The half power beam width of the telescope at 21cm is 3.6'(EW)x22'(NS)
at zero declination.  The spectrometer is a 1024-channel autocorrelator of
6.4 MHz bandwidth, with 512 channels in each polarization. The channel 
resolution is 2.64 km s$^{-1}$ at 21cm. 
Three runs (a total of 281 hours) were allocated 
for the barred subsample in 1993-94, and six runs (214 hours) for the 
unbarred subsample in 1996-99.
We observed a total of 111 galaxies, and detected 80 (+ one companion),
of which 64 are new detections.

The data were reduced at Nan\c cay Observatory with the DAC and
SIR softwares, operating on a microVax.  The horizontally and vertically
polarized signals were reduced separately and added after baseline
subtraction.  The latter operation was generally done using a polynomial of 
order 3.  All spectra were then boxcar smoothed to a final resolution of 
7.92 km s$^{-1}$. 

There was no independent check of the absolute flux calibration; 
we relied on the results of Theureau et al. (1998, their Fig.4) 
and applied a correction of 1.12 to the fluxes obtained in the
first series of runs, and of 1.32 for the second, to make up for 
successive changes in overall gain at zero declination and redshift.  
A chromatic correction
was likewise applied using a series of calibrations provided
by Gilles Theureau, where the temperature of the noise diode is seen 
to increase with decreasing frequency (increasing redshift);
this correction is similar to that adopted by Thuan et al. (1999),
but extends further in frequency.
The gain of the antenna is 1.1 K/Jy at zero declination. 
A correction for loss of sensitivity with declination
was done using graphs provided by the telescope staff.

The observations at a recession velocity of about 12500 km s$^{-1}$ were 
considerably perturbed by radar signals, and some of the galaxy profiles
around that velocity might not be reliable.  In particular,
we had to discard the spectra of Mrk 126, 413, 726,
1231.  We also discarded the spectra of Mrk 572 and 904, because
of obvious artefacts in the profiles or baselines.  This leaves 106
galaxies, out of which 24 were not detected, and three (Mrk 333, 445, 
and UGCA 304) are not part of the sample.
Mrk 445 was observed by mistake (instead of Mrk 446), 
Mrk 333 turned out to be undetected by IRAS, and UGCA 304 is a
companion to Mrk 1337.  UGCA 304 was probably not completely in
the beam of the telescope, and the corresponding Nan\c cay HI flux 
is thus underestimated.  In summary, the total number of Mrk IRAS galaxies
detected by us is 80.

In order to determine the total HI fluxes and linewidths we
converted the line profiles into MIDAS tables and images and 
applied custom-made procedures to derive these astrophysical quantities.
The total HI fluxes were corrected for beam-filling
using the precepts of Theureau et al. (1998) and taking galaxy position
angles and diameters in LEDA; they amount to a factor of 
1.02 or less in most cases.  The systemic velocity is given
in the radio convention (namely c($\nu - \nu_o$)/$\nu_o$).
At this stage, the linewidths are corrected for cosmological stretching,
but not for instrumental resolution or internal velocity dispersion.
These corrections (e.g. Fouqu\'e et al. 1990) 
will be applied before computing the maximum velocity of rotation.

For undetected galaxies, we determined an upper limit to the flux by
measuring the rms noise and estimating the linewidth 
from the absolute magnitude and inclination of the galaxy via the 
TF relation. The HI profiles are displayed in Figs.~\ref{hiprof1} 
to \ref{hiprof3}.

The uncertainties in the velocities and fluxes were estimated following
the method outlined in Fouqu\'e et al. (1990).
\begin{equation}
\Delta V_{\odot} =  4\sqrt{RP}/SNR
\end{equation}
\noindent
and
\begin{equation}
\Delta F_{\rm HI} =  5\sqrt{RAh}/SNR
\end{equation}
\noindent

\noindent
where $R$ is the channel resolution (identical to the channel spacing),
$P$ is the steepness of the profile (i.e. ($W20$ -- $W50$)/2), 
$h$ is the peak intensity in the profile (in Jy),
$A$ is the measured area under the profile (in Jy km s$^{-1}$)
and $SNR$ the signal-to-noise ratio (defined as the ratio of $h$ to the
rms noise).
The uncertainties on the linewidths $W50$ and $W20$ are simply 2 and 3 times
$\Delta V_{\odot}$ respectively.

\section{Results}
\label{results}

The results are presented in Table~\ref{Tdata}.
The Mrk number is in col.\ 1, the morphological type $t$ in col.\ 2, the
inclination in col.\ 3,
the radial heliocentric HI velocity 
(in brackets when from LEDA) and its error (in \kms) in cols.\ 4 and 5,
the rms noise in the profile (in mJy) in col.\ 6, the signal-to-noise
ratio (SNR) in col.\ 7,
the measured linewidths at 20 and 50\% of maximum intensity
(in \kms) in cols.\ 8 and 9. When no line was detected, the
listed W50 was estimated from the TF relation
and is given in brackets. 
The HI flux \fhi\ (or an upper limit) and its uncertainty (in \jykms) 
are given in cols.\ 10 and 11.
The last galaxy of that Table is UGCA 304, the companion of Mrk 1337.

The mean value of $SNR$ (col.7) is 5.5 $\pm$ 0.3, while the mean 
signal-to-noise
ratio for the fluxes (col. 10) is 4.4 $\pm$ 0.3.  The uncertainty
estimate for the HI fluxes is thus rather conservative.

HI data for 21 of the galaxies selected for observation
appeared in the LEDA database after our
observing runs were initiated. A comparison with our own data is
given in Table~\ref{Tcomp}.  The Mrk number is in col.\ 1,
the radial heliocentric HI velocity 
(in brackets when from LEDA) and its uncertainty (in \kms) in cols.\ 2 and 3,
the measured linewidths at 20 and 50\% of maximum intensity
and their uncertainty (in \kms) in cols.\ 4 to 7. 
An asterisk (*) following $W20$ means that this is in fact $W25$.
The maximum velocity of rotation and its uncertainty in log(\kms) are given
in col.\ 8 and 9,
the HI flux \fhi\ (or an upper limit) and its uncertainty (in \jykms) 
are given in cols.\ 10 and 11. The signal-to-noise in col.\ 12 and
the reference to  the measurement in col\ 13.
The references are coded as follows.
1 : this paper; 2 : Theureau et al., 1998; 3 : Giovanelli \& Haynes, 1993;
4 : Haynes \& Giovanelli, 1991; 
5 : Henning, 1992; 6 : Davis \& Seaquist, 1983; 7 : Bushouse, 1987;
8 : Giovanelli \& Haynes, 1985;
9: Andreani et al., 1995 10: Thuan et al., 1999; 11: Smoker et al., 2000;
12: Garcia et al., 1994; 13: Heckman et al., 1978;
14: Haynes et al., 1997.  

The maximum velocity of rotation is defined as (Theureau et al. 1998) :

\begin{equation}
\log(V_M)  =  (2\log(W20_c) + \log(W50_c))/3 -\log(2{\rm sin}(incl))
\end{equation}
\label{eqlvm}

\noindent
where $W20_c$ and $W50_c$ are the linewidths corrected for instrumental
resolution and internal velocity dispersion.
The upper limits to the two HI fluxes taken from the literature were
rederived using the linewidths determined by us.

The overall agreement between our data and the literature is satisfactory.
The only discrepancy in the linewidths and \lvm concerns Mrk 1199.
(Note that the discrepancy in the \lvm of Mrk 1200 is obviously 
due to the choice of inclination, which is not specified in ref. 2).
The agreement among total HI fluxes is not as good, but when 3 sets
of independent observations are available it appears that the
fluxes of ref. 2 (Theureau et al. 1998) tend to be significantly lower
than the other two. In the case of Mrk 860, the accuracy of the
flux and error-bar quoted in the literature are questionable, in view
of the rather short integration time.  
We are thus confident in the quality of our
data and adopt them in all conflicting cases.

\begin{table*}
\caption[]{Comparison of velocities, line-widths and fluxes 
from our observations and the literature
}
{\scriptsize 
\begin{flushleft}
\begin{tabular}{rrrrrrrrrrrrr}
\noalign{\smallskip}
\hline
\hline
\noalign{\smallskip}
Mrk&$V_{\odot}$&$\Delta V_{\odot}$ 
&$W_{\rm 20}$&$\Delta W_{\rm 20}$
&$W_{\rm 50}$&$\Delta W_{\rm 50}$
&$log(V_M)$&$\Delta log(V_M)$ 
&$F_{\rm HI}$&$\Delta F_{\rm HI}$&SNR&ref.\\
\noalign{\smallskip}
(1)&(2)&(3)&(4)&(5)&(6)&(7)&(8)&(9)&(10)&(11)&(12)&(13)\\
\noalign{\smallskip}
\hline
  84& 6098&18&310&54&158&36&2.119&0.120& 4.30&0.99&5.0& 1\\
  84& 6108&11&247&32&231&21&2.058&0.066& 2.7 &0.7 &4.3& 2\\
\noalign{\smallskip}
  90& 4253&13&264&39&221&26&2.390&0.044& 2.74&0.78&4.0& 1\\
  90& 4252& 7&249&21&234&14&2.369&0.025& 1.8 &0.4 &6.6& 2\\
\noalign{\smallskip}
 271& 7541&23&340&69&243&46&2.339&0.088& 2.41&0.74&3.4& 1\\
 271& 7546&  &282*& &   &  &2.323&& 1.71&    &   & 7\\
\noalign{\smallskip}
 306& 5609& 8&257&24&176&16&2.106&0.052& 7.89&0.96&8.9& 1\\
 306& 5605&  &246&  &153&  &2.074&& 5.42&    &30.0& 3\\
 306& 5778&  &594*& &   &  &2.554&&11.90&    &   & 7\\
\noalign{\smallskip}
 359&[5043]& &   &  &162&  &&&$<$2.19&    &   & 1\\
 359& 5078&  &203&  & 68&  &1.933&0.114& 0.66&    &7. & 3\\
\noalign{\smallskip}
 374&12780& 8&121&24& 74&16&1.673&0.142& 8.54&1.94&6.9 &1\\
 374&     &  &   &  &   &  &&&$<$0.29&    &   &13\\
\noalign{\smallskip}
 382&10099&16&189&48&149&32&2.285&0.069& 1.15&0.59&3.2& 1\\
 382&     &  &   &  &   &  &&&$<$0.99&    &   &13\\
\noalign{\smallskip}
 592& 7716& 5&181&15&167&10&2.022&0.040& 5.73&1.39&5.1& 1\\
 592& 7735&  &250&  &219&  &2.161&& 4.10&    &7.3& 4\\
 592& 7736& 8&   &  &175& 8&1.972&& 2.5 &0.5 &   &11\\
\noalign{\smallskip}
 860& 6868&10&140&30&104&20&1.811&0.129& 1.50&0.52&4.6& 1\\
 860& 6881&10&   &  &100&30&1.615&0.202& 0.25&0.1 &   & 9\\
\noalign{\smallskip}
 898& 5072&13&296&39&256&26&2.224&0.065& 1.94&0.63&4.0& 1\\
 898& 5086&10&274&31&267&21&2.166&0.083& 1.5 &0.5 &3.4& 2\\
\noalign{\smallskip}
 922& 6574&10&145&30& 95&20&&& 3.18&0.99& 5.4& 1\\
 922& 6576&  &143&  &111&  &&& 3.04&0.45&14.9& 8\\
\noalign{\smallskip}
 929& 5259&16&296&48&223&32&2.232&0.078& 4.43&1.04& 4.3& 1\\
 929&     &  &278&  &   &  &2.258&& 3.76&    &     &8\\
\noalign{\smallskip}
1026& 3994& 8&271&24&205&16&2.221&0.040&12.87&1.59& 8.2& 1\\
1026& 3989& 6&260&19&212&13&2.151&0.075& 7.7 &0.9 &11.7& 2\\
1026& 3961&15&274&45&201&30&2.221&0.075& 8.3 &1.9 &    &12\\
\noalign{\smallskip}
1039& 2111& 5&161&15&134&10&1.860&0.057& 6.16&0.97& 8.3& 1\\
1039& 2098&65&198&17&149&11&1.940&0.623& 6.76&0.40&    & 10\\
\noalign{\smallskip}
1092& 3356& 3&338&9&311&6&2.215&0.015&19.90&1.22&14.7& 1\\
1092& 3363& 4&336&13&319& 9&2.174&0.020&12.2 &1.3 &10.8& 2\\
\noalign{\smallskip}
1104& 1991&37&155&111&109&74&2.139&0.224&16.85&4.99& 1.4& 1\\
1104& 1985&  &135&  &107&  &2.095&&21.12&    &    & 5\\
1104&     &  &197*& &   &  &2.335&&14.60&    &     &6\\
\noalign{\smallskip}
1198& 7635&10&509&30&466&20&2.414&0.032& 9.63&1.79& 4.9& 1\\
1198& 7647&12&511&35&495&23&2.391&0.031& 3.4&0.8 & 4.0& 2\\
1198&     &  &   &  &   &  &&& 7.2&0.36&    &14\\
\noalign{\smallskip}
1199& 4085&17&171&51& 60&34&1.859&0.196& 1.78&0.67& 4.9& 1\\
1199& 4031&14&282&43&170&29&2.165&0.117& 3.1 &0.6 & 7.5& 2\\
\noalign{\smallskip}
1200& 7788& 8&293&24&253&16&2.486&0.022& 4.45&0.77& 6.2& 1\\
1200& 7784&14&248&42&242&28&2.928&0.014& 0.9 &0.5 & 2.4& 2\\
\noalign{\smallskip}
1270& 8307&14&299&42&196&28&2.250&0.066& 7.60&1.41& 5.8& 1\\
1270& 8357&18&314&55&187&37&2.219&0.141& 3.0 &0.6 & 6.2& 2\\
\noalign{\smallskip}
1363& 2743& 3&206&9&197&6&2.236&0.015& 9.46&1.09& 9.9& 1\\
1363& 2736& 3&206&10&200& 7&2.239&0.014& 5.8 & .9 &10.2& 2\\
\hline
\hline
\end{tabular}
\end{flushleft}

}
\label{Tcomp}
\end{table*} 

We next examine the individual profiles, taking into account the morphological
type and inclinations given in Table~\ref{Tdata}.
Only a few of the galaxies of our sample have the standard saddle-shaped
profile: Mrk 1092, 1337, 1363, 1425 and, to a lesser extent, 
Mrk 90, 382, 592, 827, 898, 905, 966, 968, 997, 1200, 1270.
Only one narrow profile corresponds to a nearly face-on galaxy (Mrk 626); 
the others (Mrk 373, 922, 928, 937, 1064, 1104) suggest that the HI may
in some instances be confined to the central parts of the galaxy.
Such a behavior has also been noted for the molecular gas of these
starburst galaxies (Contini et al. 1997a).
Conversely, some nearly face-on galaxies, like Mrk 922 and 1200, have 
large line widths, suggesting non-planar motions.
Most profiles appear perturbed, but,
because of the generally low signal-to-noise ratio, it is
difficult to draw more information about the dynamics of the
neutral gas from these profiles.
 
Finally, we note that our data on Mrk 237 might be affected by 
confusion, since the galaxy has a very close companion at about 
the same recession velocity.

\section{Analysis}

We complemented the HI data from our 80 galaxies by data 
from the literature for 138 other galaxies, to form a large
sample of 218 Mrk IRAS galaxies, 105 unbarred and 113 barred ones, 
which is given in Table~\ref{Tdera}. The Mrk number is in col.\ 1, the
morphological type (SA = unbarred, SB = barred) in col.\ 2, 
the spectral type (stb = nuclear starburst; sy = Seyfert; st? and sy? are
classifications from an IRAS color index as explained below) in col.\ 3,
the numerical morphological type $t$ (defined in RC3;
de Vaucouleurs et al. 1991) in col.\ 4,
the inclination (in degrees) in col.\ 5,
the isophotal diameter $d_{\rm c}$ and its uncertainty (in log of 0.1\arcmin) 
in cols.\ 6 and 7, the maximum rotation velocity \vm\ (in \kms) and 
its uncertainty in cols.\ 8 and 9, the distance $D$ (in Mpc) in
col.\ 10, the $K$-band magnitude in col.\ 11, the oxygen abundance  (O/H)
with respect to solar in col.\ 12, the total HI mass and its uncertainty
in \msun\ in cols.\ 13 and 14, the total far-infrared (FIR) luminosity 
(in \lsun\ ) in col.\ 15 and the absolute blue magnitude \mabs\ in col.\ 16.

The spectral type (starburst or Seyfert) was taken from Contini et al.
(1998), Mazzarella \& Balzano (1986) or Mazzarella \& Boroson (1993). 
When no classification was available, we estimated it from the
IRAS color index $\alpha$(60,25) = $\log(S_{60}/S_{25})$ / $\log(25/60)$, 
where $S_{60}$ and $S_{25}$ are the IRAS fluxes at 60 and 25 \micron\ 
respectively.
According to Coziol et al. (1998), $\alpha$(60,25) is larger than $-1.70$
for Seyferts and between $-2.5$ and $-1.70$ for starburst galaxies; this last
spectral classification is of course to be taken with caution.
The maximum rotation velocity \lvm was computed from the linewidths
(given in Table~\ref{Tdata}) corrected for resolution and internal 
velocity dispersion (eq.~\ref{eqlvm}),
or taken from LEDA when not in Table~\ref{Tdata}.

\begin{figure}[t]
\centering
\includegraphics[width=87mm]{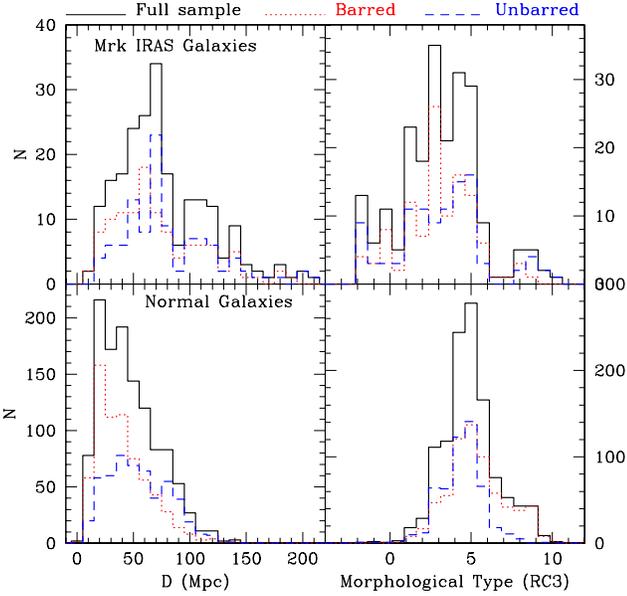}
\caption{Distribution of distances ({\it left}) and morphological
types ({\it right}) for the sample of Mrk IRAS galaxies ({\it top}), and 
for a sample of normal galaxies from Mathewson \& Ford (1996) ({\it bottom}). 
A distinction is made between barred (dotted line) and unbarred (dashed line)
galaxies.  The Mrk IRAS galaxies are on average further away and of earlier 
morphological type than the normal galaxies 
}
\label{histodist}
\end{figure}

\begin{figure}[t]
\centering
\includegraphics[width=87mm]{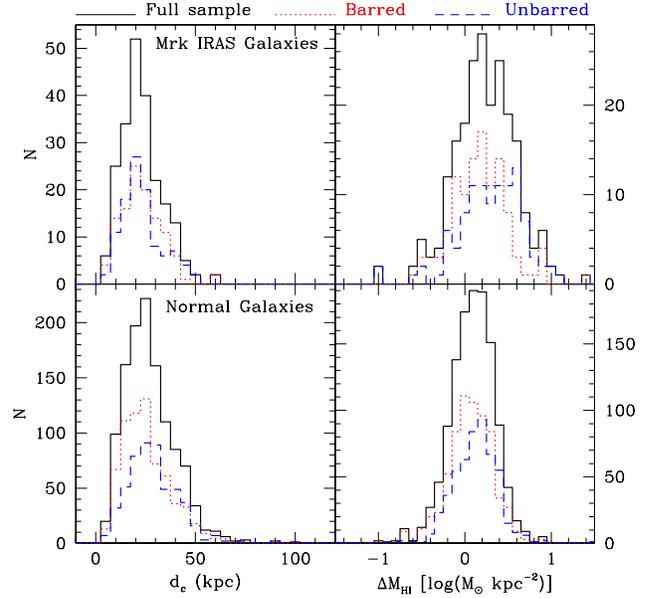}
\caption{Distribution of isophotal linear diameters ({\it left})
and excess of ({\it right})
HI mass for the sample of Mrk IRAS galaxies ({\it top}), and for a sample of 
normal galaxies from Mathewson \& Ford (1996) ({\it bottom}). 
A distinction is made between barred (dotted line) and unbarred (dashed line)
galaxies.  The Mrk IRAS galaxies have on average a larger HI mass excess
than the normal ones. The barred galaxies have on average
a smaller HI mass excess
than the unbarred ones; this is a property of spiral galaxies in general 
}
\label{histomhi}
\end{figure}

The distance was estimated from 
the recession velocity corrected for Virgocentric flow (given in LEDA)
and a Hubble constant of 75 \kmsmpc.  
The $K$-band magnitude was taken from NED\footnote{NED 
(http://nedwww.ipac.caltech.edu)} or from 
Bergougnan et al.\ (paper VII, in preparation).
The oxygen abundance was taken from Contini (1996) for the barred (SB) 
galaxies and from the literature, mainly Balzano (1983) and Dahari \& 
De Robertis (1988), for the unbarred (SA) ones. The O/H for the SA are 
generally based on photographic work and thus more uncertain than that 
for the SB.
The total HI mass was derived from the 
corrected total fluxes (given in Table 1) according to the following equation 
(Haynes \& Giovanelli 1984):
\begin{equation}
\log(M_{\rm HI}) =  5.373 + 2\log(D) + \log(F_{\rm HI})
\end{equation}
For galaxies not observed at Nan\c cay, it was estimated
from the $m21_{\rm c}$ parameter in LEDA using the equation :
\begin{equation}
\log(M_{\rm HI}) =  12.335 + 2\log(D) - 0.4m21_{\rm c}
\end{equation}

The morphological type, inclination, isophotal diameter, FIR luminosity
and absolute magnitude were taken from LEDA.

\subsection{Neutral hydrogen content}
\label{hicontent}

\begin{table*}
\caption{Mean values and uncertainties of the morphological type $t$,
distance $D$, linear diameter $d_c$, HI mass $M_{HI}$, HI mass to blue
light ratio $M_{\rm HI}/L_{\rm B}$, HI mass density $M_{\rm HI}/d_{\rm c}^2$ and
HI mass excess $\Delta M_{\rm HI}$, for the sample of 
Mrk IRAS galaxies and for the comparison sample of 
normal galaxies (Mathewson \& Ford 1996).}
\begin{center}
\begin{tabular}[c]{lrccccccl}
\hline
\noalign{\smallskip}
Sample & $N$ & $t$ & $D$ & $d_{\rm c}$ & $\log(M_{\rm HI})$ &
$\log(M_{\rm HI}/L_{\rm B})$  &
$\log(M_{\rm HI}/d_{\rm c}^2)$  & $\Delta M_{\rm HI}$\\
\noalign{\smallskip}
 & & & [Mpc] & [kpc] & [log(\msun)] & 
[log(\msun\lsun$^{-1}$)] &
[log(\msun kpc$^{-2}$)] & [log(\msun kpc$^{-2}$)] \\
\noalign{\smallskip}
\hline
\noalign{\smallskip}
\multicolumn{9}{c}{\sl Mrk IRAS Galaxies}\\
\noalign{\smallskip}
All     &218&3.08$\pm$0.17 & 76.6$\pm$2.8 & 23.3$\pm$0.7 & 9.69$\pm$0.03 & 
-0.63$\pm$0.03&7.04$\pm$0.03 & 0.24$\pm$0.03 ($N$=199)\\
Barred  &113&3.04$\pm$0.21 & 71.9$\pm$3.7 & 23.5$\pm$1.0 & 9.65$\pm$0.04 & 
-0.65$\pm$0.03&6.99$\pm$0.04 & 0.20$\pm$0.03 ($N$=106)\\
Unbarred&105&3.13$\pm$0.27 & 81.7$\pm$4.0 & 23.1$\pm$1.0 & 9.73$\pm$0.04 & 
-0.60$\pm$0.03&7.09$\pm$0.04 & 0.29$\pm$0.04 ($N$=93)\\
\noalign{\smallskip}
\multicolumn{9}{c}{\sl Normal Galaxies (Mathewson \& Ford 1996)} \\
\noalign{\smallskip}
All     &1197&4.92$\pm$0.05 & 46.5$\pm$0.7 & 27.0$\pm$0.4 & 9.70$\pm$0.01 & 
-0.52$\pm$0.01&6.93$\pm$0.01 & 0.10$\pm$0.01 ($N$=1194)\\
Barred  & 678&5.31$\pm$0.07 & 40.5$\pm$0.9 & 25.3$\pm$0.5 & 9.62$\pm$0.02 & 
-0.54$\pm$0.01&6.91$\pm$0.01 & 0.07$\pm$0.01 ($N$=677)\\
Unbarred& 519&4.41$\pm$0.06 & 54.2$\pm$1.2 & 29.2$\pm$0.6 & 9.80$\pm$0.02 & 
-0.50$\pm$0.02&6.95$\pm$0.01 & 0.13$\pm$0.01 ($N$=517)\\
\hline
\end{tabular}
\end{center}
\label{meandata}
\end{table*}

The first questions to address are {\sl i)} whether Mrk IRAS galaxies have an
abnormal neutral hydrogen content with respect to normal galaxies, and
{\sl ii)} whether the HI content of barred and unbarred galaxies
differ.  As comparison sample of ``normal'' galaxies, we used the 
the sample of Mathewson \& Ford (1996).
In order to compare homogeneous sets of data, we only selected the 1197 normal
galaxies for which the maximum velocity and HI flux were available in LEDA. 

The Mrk IRAS galaxies have on average the same quantity of HI mass as the
normal galaxies (see Table~\ref{meandata}).
However, this result is probably affected by selection effects.
Indeed, our sample of Mrk IRAS galaxies and that 
of normal galaxies do not have the same distributions in distance, 
nor in morphological type
(see Fig.~\ref{histodist}); the Mrk IRAS galaxies are further away and 
of earlier type than the normal galaxies.  Galaxies observed further
away tend to have higher masses because of the Malmquist bias, and
later-type galaxies tend to be richer in HI.

In order to circumvent the distance bias, we normalize 
the HI mass by $d_{\rm c}$ (the corrected linear diameter at the isophote 
25 blue mag arcsec$^{-2}$) squared, which is free of 
Malmquist bias.  This has the dimension of an HI surface density;
but one has to keep in mind that the blue isophotal diameter does not 
necessarily measure the radial extent of the HI distribution. The
distribution of isophotal diameters $d_{\rm c}$ is shown in
Fig.~\ref{histomhi}, for the sample of Mrk IRAS 
galaxies and for the comparison sample of normal galaxies. 
Mean values of isophotal diameters and HI surface densities, 
$M_{\rm HI}/d_{\rm c}^2$, with uncertainties are listed 
in Table~\ref{meandata},
where a distinction is made between barred and unbarred galaxies.

We find that, on average, the Mrk IRAS have a higher surface density
than the normal galaxies. But, as noted above, the two samples do not have 
the same distributions in morphological types, which introduces another bias. 
To take these differences into account,
we calculate the excess HI mass over that predicted
for a given morphological type and diameter ($d_c$, in kpc),
using the following equation :

\begin{equation}
\Delta M_{\rm HI} =  \log(M_{\rm HI}) - c_1 -c_2\log(d_c^2)
\end{equation}

\noindent
where the constants $c_1$ and $c_2$ are given in Table V of 
Haynes \& Giovanelli (1984)
for the different morphological types, except those earlier than S0
($t=-1.5$).
The values of excess HI mass for the different subsamples are listed
in Table~\ref{meandata}, together with the number of objects 
(since types earlier than S0 were not included).  
As a check, we corrected the HI surface densities 
in another way, using as reference the mean HI surface densities 
given by Giovanelli \& Haynes (1988; their Table 12.1);
this method leads to results that are in qualitative agreement with
those obtained by the method of Haynes \& Giovanelli (1984).

We find that Mrk IRAS galaxies have a larger excess ($\Delta M_{HI}$) 
of HI mass than normal galaxies.
The fact that most galaxies have a positive excess of HI mass is 
due to the definition of the diameter: Haynes \& Giovanelli (1984) 
use the uncorrected diameter from UGC, whereas we use the corrected 
diameter from LEDA, and the LEDA uncorrected angular diameters are
on average 9\% smaller than the UGC diameters
(Paturel et al. 1991).

We also computed the HI mass to blue light ratio for the two samples,
and find that the Mrk IRAS galaxies have a lower ratio than the normal
galaxies. This is contrary to expectations, and probably due to the
fact that the two samples have different distributions in type 
(see Fig.~\ref{histodist}) : the Mrk IRAS are of earlier type, thus
of brighter absolute magnitude (see Roberts \& Haynes 1994) than
the other sample. There is in fact a small magnitude dependence
in the HI mass to blue luminosity ratio expressed as 
$L_{\rm B}^{\beta}$ (see Smoker et al. 2000
and references therein). We find that $\beta$ = -- 0.33 for
both our samples.  Computing $M_{\rm HI} L_{\rm B}^{-0.66}$ reduces,
but does not reverse the trend.  But this is perhaps not a very
good HI mass indicator, since the exponent $\beta$ 
depends on the sample, and ranges from -- 0.1 to -- 0.33.

We finally computed the excess HI 
mass over that predicted for a given morphological type and luminosity, 
using the constants given in Table V of Haynes \& Giovanelli (1984).
For this type of bias correction we find the expected trend,
that the Mrk IRAS galaxies have on average a larger HI mass excess
than the normal galaxies, and that the normal barred galaxies have
less excess than the normal unbarred galaxies (there is no significant
difference between the barred and unbarred Mrk IRAS galaxies in that
respect).  

These last results weaken our
claim that the Mrk IRAS galaxies have more HI than normal galaxies.
However, as stated by Haynes \& Giovanelli (1984), normalizing the
HI mass by the luminosity leads to HI mass indicators of inferior
quality compared to $M_{\rm HI}/d_{\rm c}^2$. 

The answer to the second question, whether barred and unbarred galaxies
have different HI content, is perhaps easier to give, since 
the detection of a bar is less likely to be biased. One could argue
that bars are less conspicuous in more distant galaxies (and, indeed,
the unbarred galaxies tend to be further away than the barred ones
in Table~\ref{meandata}), but
our classification (in the Mrk IRAS sample) is based on high resolution
images (Contini, 1996), thus less affected by the Malmquist effect. 

\begin{table}
\caption{Kolmogorov-Smirnov test on the significance of the
differences between samples for various parameters. The first
three parameters are tested in logarithms. The numbers are confidence
levels (in percent) that the two samples are drawn for different
populations.
}
\begin{center}
\begin{tabular}[c]{lllll}
\hline
\noalign{\smallskip}
Sample &\multicolumn{4}{c}{Parameter}\\
\noalign{\smallskip}
& $M_{\rm HI}$ & $M_{\rm HI}/L_{\rm B}$  &
$M_{\rm HI}/d_{\rm c}^2$  & $\Delta M_{\rm HI}$\\
\noalign{\smallskip}
\hline
\noalign{\smallskip}
Mrk {\sl vs} Normal & 46.8 & 99.9 & 99.9 & 99.8\\
Mrk : SA {\sl vs} SB& 90.7 & 69.6 & 95.5 & 99.1\\
Normal : SA {\sl vs} SB& 99.9 & 97.0 & 98.0 & 99.9\\
\noalign{\smallskip}
\hline
\end{tabular}
\end{center}
\label{TKS}
\end{table}

We find that barred galaxies have on average a lower HI mass,
lower HI mass to blue luminosity,
lower HI surface density and lower HI mass excess than unbarred ones 
(see Table~\ref{meandata}),
and this is true for both Mrk IRAS and normal galaxies.  

We performed Kolmogorov-Smirnov tests to quantify the statistical
significance of the differences found between samples. The results
are summarized in Table~\ref{TKS}, which gives the confidence level
(in percent) at which the two samples are drawn for different
populations.
The difference in the parameters $\log(M_{\rm HI}/d_{\rm c}^2)$ and 
$\Delta M_{\rm HI}$ is significant at a confidence level of at least 95\%,
and mostly of 98-99\%.  As stated above, the significance of the differences
in $\log(M_{\rm HI})$ is probably illusory in view of possible Malmquist
bias.

In summary, the two main results of this subsection
are that {\sl i)} Mrk IRAS galaxies 
have a larger excess of HI mass than normal galaxies, and {\sl ii)}, 
within each category (Mrk IRAS or normal), unbarred galaxies have 
a larger HI mass, mass density and mass excess than barred ones.
This is a clear indication that activity (starbursts or Seyfert) and 
the presence of a bar are both related to the HI content of galaxies.

Finally, we must point out that we did not find any dependence of the 
HI mass excess on the global star formation rate of galaxies, measured 
by the FIR luminosity.  This also holds for barred and unbarred
galaxies considered separately, and by excluding Seyferts and uncertain
types of nuclear activity (st?, sy?; see Table~\ref{Tdera}). 
On the other hand, we do find the correlation between
star formation rate and HI surface density which has been established for
different samples of
nearby galaxies (e.g.\ Kennicutt 1989; Donas et al.\ 1990; 
Deharveng et al.\ 1994; Boselli 1994; Kennicutt 1998).  

\subsection{Tully-Fisher relation}
\label{tfrelation}

   \begin{figure*}[t]
   \centering
   \includegraphics[width=87mm]{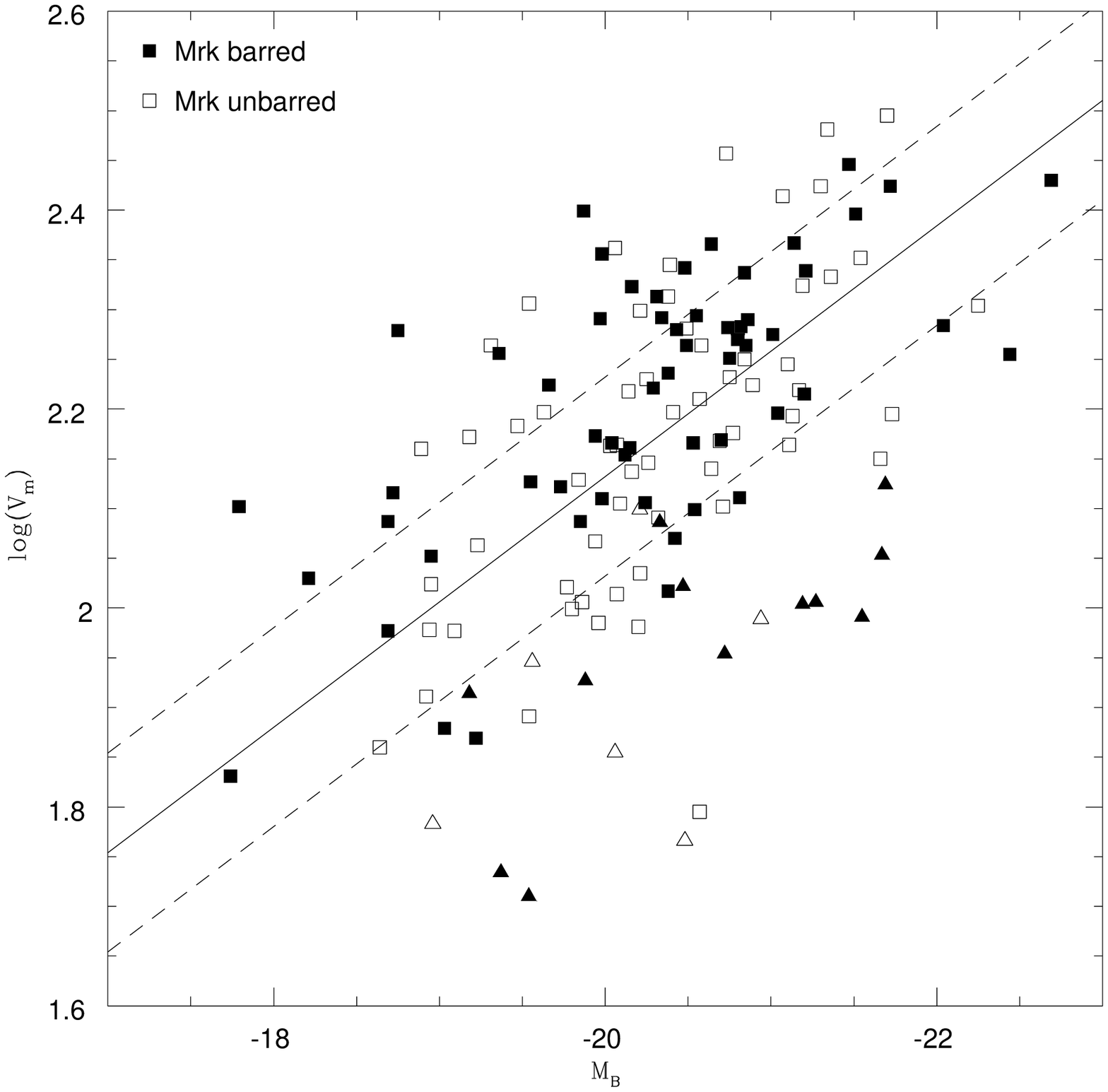}
   \includegraphics[width=87mm]{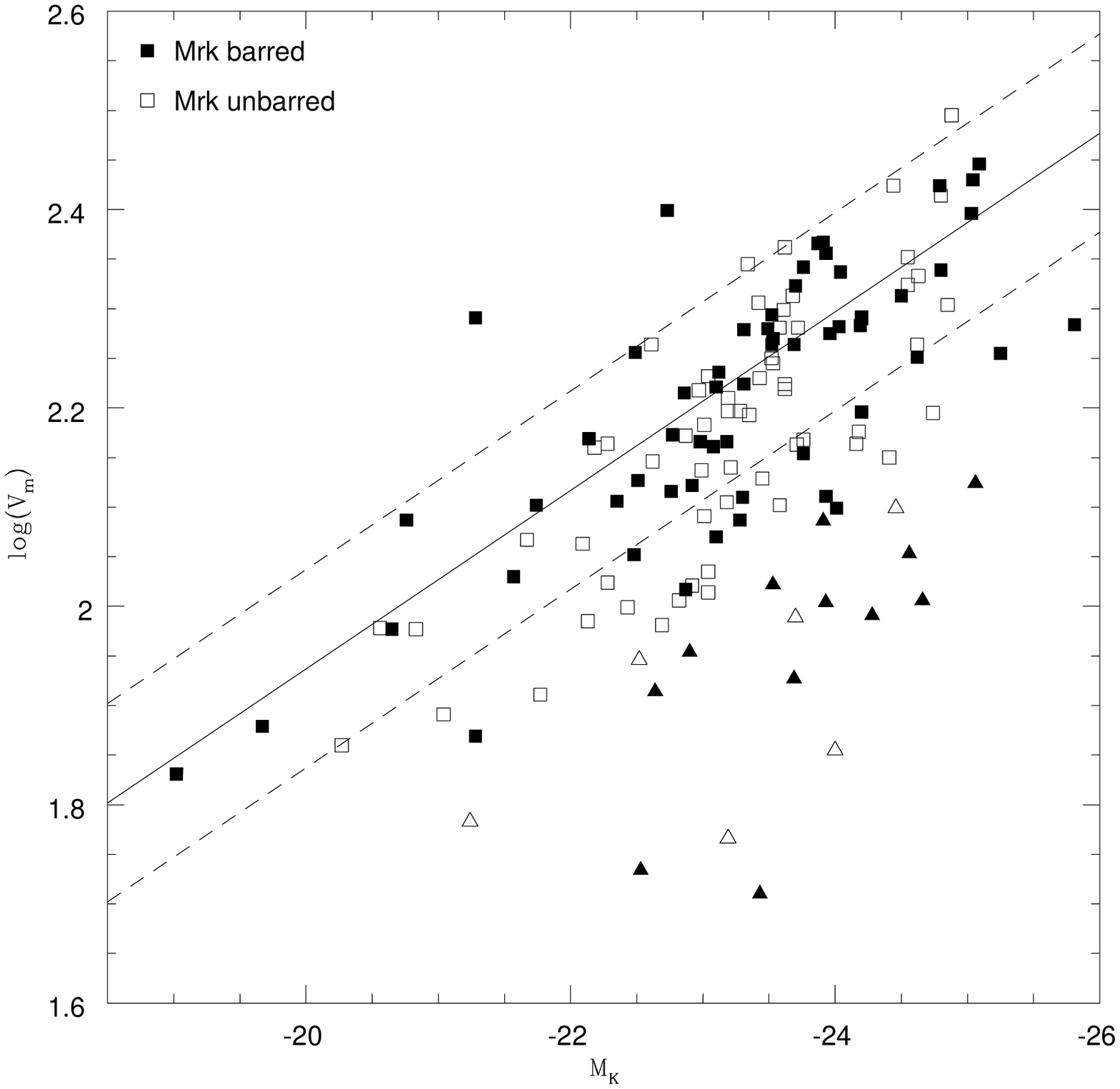}
      \caption{Tully-Fisher relation in the $B$ band ({\it left}) and 
in the $K$ band ({\it right}) for our sample of Mrk IRAS
galaxies. The full and empty squares are the barred and 
unbarred galaxies respectively. The full and empty triangles are
the barred and unbarred galaxies which deviate most from the TF
relation in the K band.  These Figures are limited to data
with an uncertainty in \lvm\ smaller than 0.1 and an inclination
larger than 30$^{\circ}$.
The solid lines are the relation for normal galaxies (Eqs.~\ref{tfb} and 
\ref{tfk}). The dashed lines delineate the zone where the data
should in principle be enclosed (uncertainty on \lvm less than 0.1).
The Mrk IRAS galaxies do not follow the local TF relations for normal
galaxies, as a significant proportion of them have a low value of \lvm
}
         \label{tfbk}
   \end{figure*}

\begin{table}
\caption{Galaxies with the strongest deviations from the $K$-band TF
relation, grouped into barred and unbarred types. 
The excess of HI mass (see sect.~\ref{hicontent}) is
given in col.\ 5. 
The difference between the observed \vm\ and that given by
Eq.~\ref{tfk} is in col.~6. 
The last column gives informations on the environment and 
on the level of interaction of galaxies (from Contini 1996; Keel \&
van Soest 1992).}
\begin{center}
\begin{tabular}[c]{rcrrrrl}
\hline
\noalign{\smallskip}
Mrk&Spe&O/H&\lfir
&$\Delta M_{\rm HI}$&$\delta\log(V_{\rm m})$&Envir. \\
\noalign{\smallskip}
(1)&(2)&(3)&(4)&(5)&(6)&(7)\\
\noalign{\smallskip}
\hline
\noalign{\smallskip}
\multicolumn{7}{c}{\sl Barred Galaxies}\\
  52&stb&0.78& 9.72&-0.14&-0.431&isolated\\
 300&stb&0.74&10.45& 0.25&-0.350&satellite\\
 307&stb&1.00&10.26& 0.22&-0.287&isolated\\
 313&sy2&    & 9.55& 0.87&-0.261&in group\\
 319&stb&0.44&10.88& 0.13&-0.295&pair (1.5\arcmin)\\
 353&stb&0.55&10.35& 0.11&-0.342&isolated\\
 471&sy2&    &10.40& 0.02&-0.269&isolated\\
 489&stb&0.57&10.74& 0.91&-0.332&isolated\\
 592&stb&    &10.36& 0.38&-0.233&isolated\\
 691&stb&1.50&10.12& 0.45&-0.244&satellite\\
1157&sy2&    &10.05&-0.07&-0.203&isolated\\
1302&stb&0.90& 9.68& 0.41&-0.535&isolated\\
mean&   &    &10.21& 0.30\\
\noalign{\smallskip}
\noalign{\smallskip}
\multicolumn{7}{c}{\sl Unbarred Galaxies}\\
\noalign{\smallskip}
 201&stb&0.44&10.59&-0.11&-0.458&merger?\\
 331&st?&    &11.08& 0.65&-0.239&pair (2.0\arcmin)\\
 685&st?&0.22& 9.65& 0.55&-0.265&isolated\\
 905&st?&    & 9.95& 0.69&-0.442&pair (1.7\arcmin)\\
1003&sy?&    & 9.36&-0.12&-0.218&isolated\\
mean&   &    &10.13& 0.33\\
\hline
\end{tabular}
\end{center}
\label{devTF}
\end{table}

The Tully-Fisher relation is one 
of the most basic relations among spiral galaxies and provides a 
critical constraint on galaxy formation theories. The third objective of this 
analysis is to examine whether the Mrk IRAS galaxies follow 
the TF relation.

We determined a TF relation for normal galaxies in the $B$ and $K$ bands 
using the sample of Mathewson \& Ford (1996), extracting the relevant 
parameters from LEDA and NED, and restricting our analysis 
to galaxies with an uncertainty smaller than
0.1 on \lvm\ and an inclination larger than 30$^{\circ}$.  
There were 1162 such galaxies for calibrating the $B$-band TF relation,
and 682 for the $K$ band. The advantage of determining our own relations,
rather than relying on relations from the literature, is that we are
comparing homogeneous sets of data, affected in the same way by
systematic effects and errors. 

\noindent
In the $B$ band, the relation is :

\begin{equation}
\log(V_{\rm m}) = -0.126 M_{\rm B} - 0.388
\label{tfb}
\end{equation}

\noindent
In the $K$ band, it is :

\begin{equation}
\log(V_{\rm m}) = -0.090 M_{\rm K} + 0.137
\label{tfk}
\end{equation}

The $B$-band TF relation for our sample of Mrk IRAS
galaxies is shown in Fig.~\ref{tfbk}, together with the
linear relation for normal galaxies. This Figure is limited 
to data with an uncertainty in \lvm\ smaller than 0.1
and inclinations larger than 30$^{\circ}$.   
This plot shows that there is no difference between the barred and
unbarred galaxies. More remarquably, it also shows that the Mrk IRAS 
galaxies do not follow the local $B$-band TF relation, as already 
stated in Coziol et al. (2000), showing a large scatter. 

One could argue that the luminosity in the 
$B$-band is dominated by young OB stars in starbursts or by the
non-thermal nucleus in Seyferts and thus is not a good tracer of the 
total virial mass.  A starburst generally brightens the galaxy
by up to 2 mag in $B$, and only 0.5 mag in $K$
(Mouhcine \& Lan\c  con 2003).  However, there is admittedly no significant
excess of galaxies with low \lvm compared to high-\lvm ones,
which is surprizing. One could perhaps argue that the B-band
TF relation estimated from normal galaxies is
affected by unidentified starburst galaxies in the 
sample, or that OB stars and dust play in
opposite senses in starburst galaxies, contributing to the observed scatter. 
At any rate, to avoid potential problems of the $B$ band, we use 
the $K$-band absolute magnitude as mass indicator.  
  
The $K$-band TF relation for our sample of starburst and Seyfert 
galaxies is shown in Fig.~\ref{tfbk}, together with the
linear relation for normal galaxies.  
This Figure is limited to data with an uncertainty in
\lvm\ smaller than 0.1 and an inclination larger than 30$^{\circ}$.  
Like the TF relation in the $B$ band, we find no difference
between the barred and unbarred galaxies, 
and a significant proportion of galaxies with 
very low \vm.  

The fact that the barred and unbarred galaxies show no difference
in the TF plane is not specific to Mrk IRAS galaxies. We have
verified that this is also true for the normal galaxies of
Mathewson \& Ford (1996). Courteau et al. (2003) reach a
similar conclusion, but based on optical kinematics.
This simply means that the presence of a bar does not
affect the proportion of luminous to dark matter at a
given radius.

We believe that the excess 
of galaxies with low \vm\ in Fig.~\ref{tfbk} is real, and tells
us something about the physical properties of Mrk IRAS galaxies.
We next try to identify the origin of this effect, by investigating
the 17 most deviant galaxies in the $K$-band TF diagram.
They are indicated by a triangular symbol in both panels of Fig.~\ref{tfbk};
one can see that most of them are also deviant in the $B$-band TF relation.
Their properties are listed in Table~\ref{devTF}.  

The HI distribution has been mapped in only two 
galaxies (Mrk 52 and 313) among those listed in Table~\ref{devTF}. 
The HI distribution of Mrk 52 is well resolved, revealing a centrally
peaked column density structure (Taylor et al. 1995). 
The velocity map of this galaxy
indicates differential rotation, though it lacks spatial and velocity
resolution to detect any asymmetries that might exist. High-resolution 
HI observations in the Mrk 313 group of galaxies reveal that the 
HI morphology of almost all the member galaxies are affected by the
environment and four galaxies show tidally distorted HI
morphology and kinematics (Li \& Seaquist 1994). 
The HI emission of the pair NGC 7464/Mrk 313 is
dominated by a ring around Mrk 313. The orientation of Mrk 313 and
the ring suggest that it represents a polar ring around Mrk 313 and
could be material pulled out of NGC 7464 during a close encounter
with Mrk 313.

For the other galaxies, we have to rely on the HI profiles. Four of them
are in Fig.~\ref{hiprof1}, the others can be found in the literature.
Only a few galaxies have a classical double-horned profile : Mrk 52, 307, 489, 
592, and possibly Mrk 300 and 319. The others have a sloping profile,
either gaussian or centrally peaked (Mrk 201, 331, 353, 691, 905, 1064, 1302)
or a lopsided profile (Mrk 313, 471, 685, 1003, 1157).  The gaussian/centrally
peaked profiles suggest that the HI is centrally concentrated and does
not extend to the flat part of the rotation curve, the lopsided profiles 
indicate uneven distribution of HI in the disk of the galaxy.
In summary, such profiles indicate that the HI gas has not 
reached equilibrium in these galaxies. 

   \begin{figure*}[t]
   \centering
   \includegraphics[width=87mm]{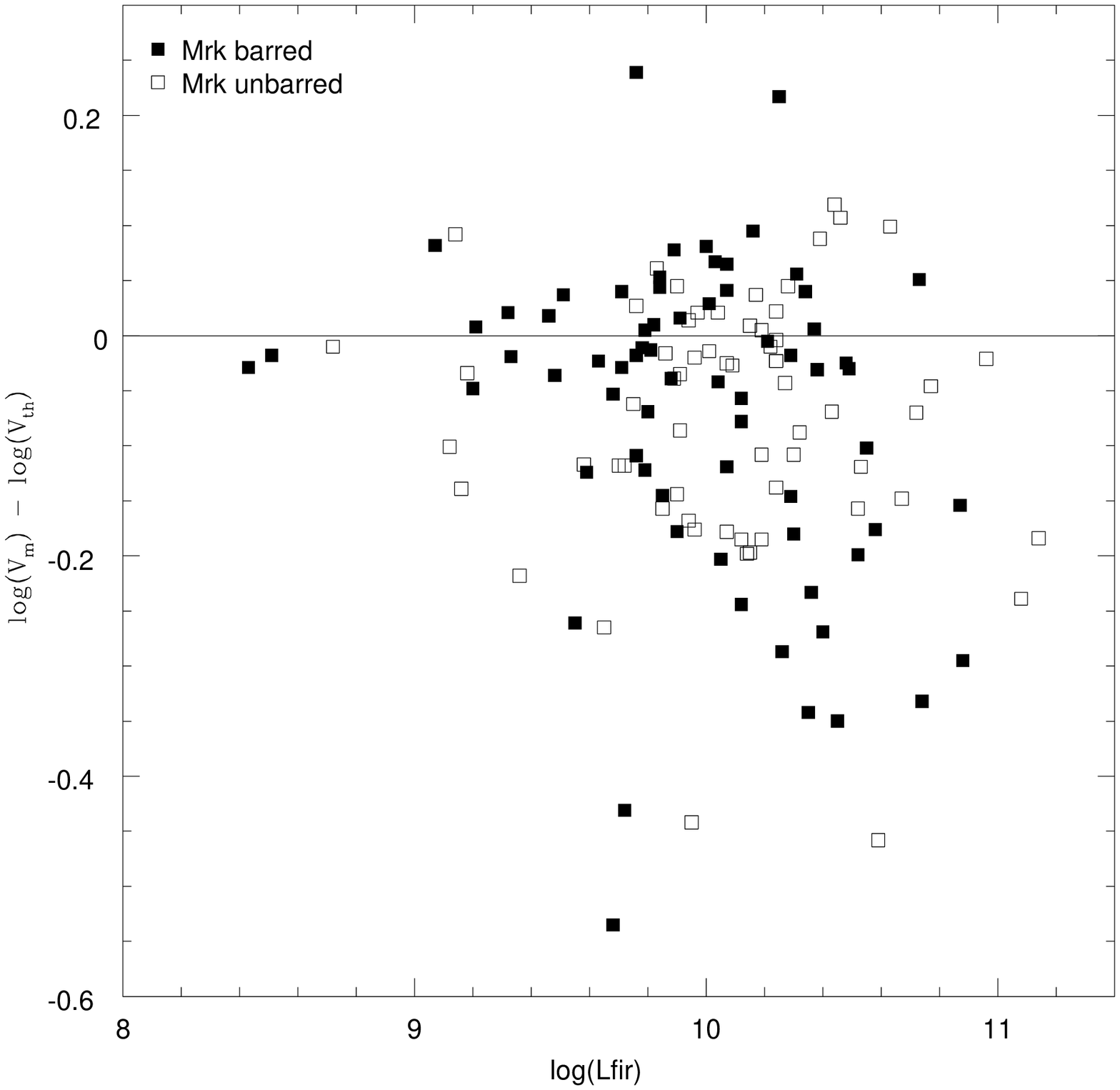}
   \includegraphics[width=87mm]{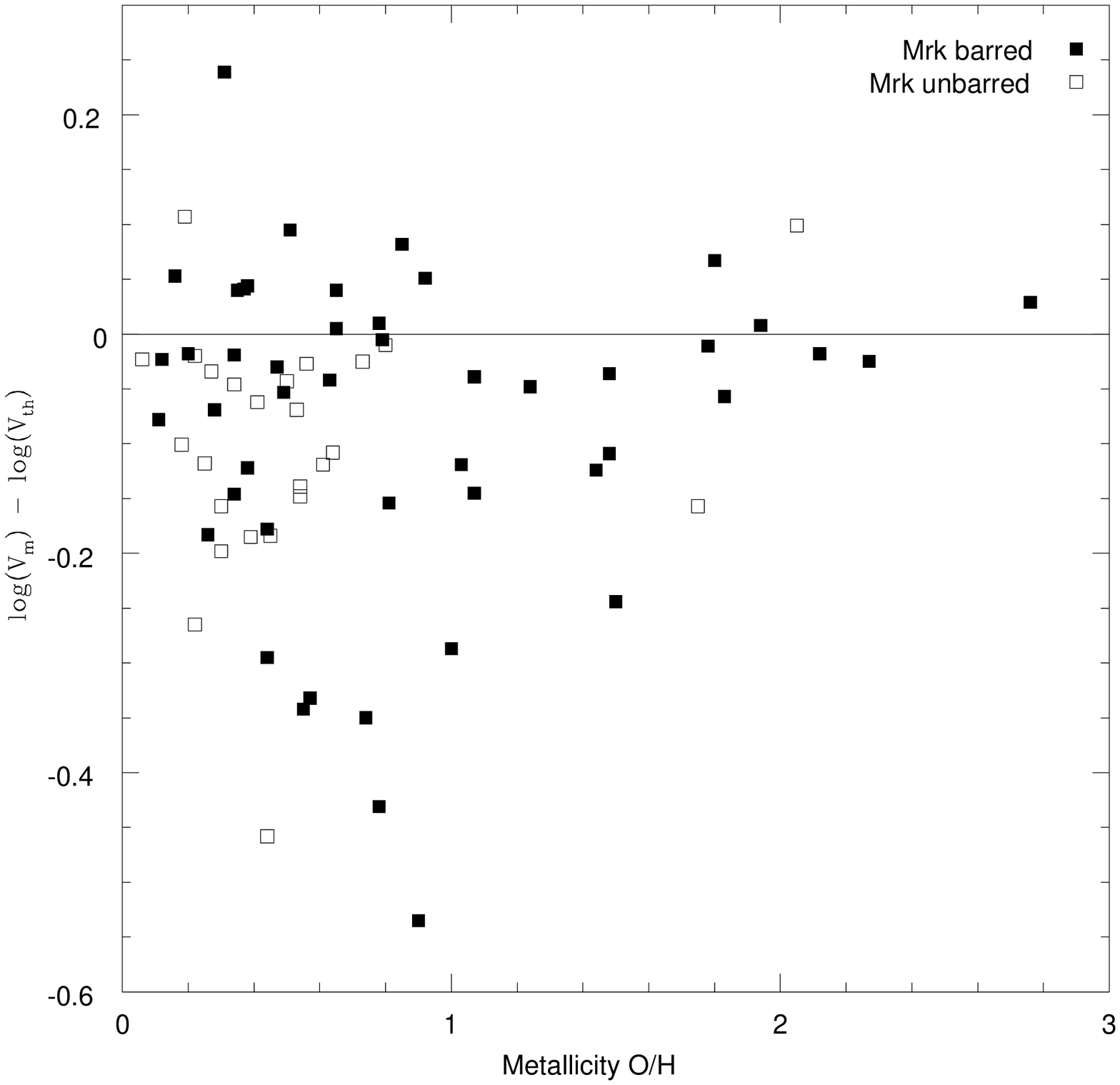}
      \caption{Departure from the TF relation in the $K$ band as 
a function of FIR luminosity ({\it left}), and oxygen abundance (O/H) 
({\it right}). Like in Fig.~\ref{tfbk}, only data points with
uncertainties in log($V_m$) less than 0.1 and inclinations higher 
than 30$^{\circ}$ are plotted.  The full and empty squares are the barred 
and unbarred galaxies respectively. 
The galaxies which depart most from the TF relation also have the 
strongest FIR luminosity and the lowest metallicity.
}
         \label{devtf}
   \end{figure*}

Peculiar HI kinematics does not seem to be limited to active galaxies. We have
examined the properties of the 7 galaxies which have an abnormally low
\lvm\ in the $K$-band TF relation for the 682 normal galaxies
(Mathewson \& Ford 1996).
Five of these galaxies (ESO 34-3, 446-2, 512-12, 563-13, 576-1)
become ``normal'' again if one uses the maximum
rotation velocity of the ionized gas instead of
that of the HI gas. For the two other ones (ESO 384-32 and UGC 645)
optical kinematics are not available. Thus peculiar HI
kinematics in an otherwise normally rotating galaxy can also be found
among normal galaxies, unless these objects are misclassified in terms
of nuclear activity.

In order to better understand the cause of the slow rotation of 
the neutral HI gas in part of our sample, we have looked for
correlations with other properties of the galaxies. 
Their environment, as determined from CCD and sky survey images, gives no
indication to explain the perturbed kinematics : only one galaxy (Mrk 201)
appears strongly disturbed, and another one (Mrk 313) is in
a group, while the others are isolated or in rather distant pairs
and fairly symmetric in shape (Table~\ref{devTF}).
The 17 deviant galaxies have on average a much higher HI mass excess 
($<\Delta M_{\rm HI}>=0.30$ and $0.33$ for the barred and unbarred 
galaxies respectively, see Table~\ref{devTF}) than
the whole sample of Mrk IRAS galaxies (Table~\ref{meandata}). 
As pointed out by the referee, applying the baryonic correction
advocated by McGaugh et al. (2000) to these gas-rich galaxies makes them
even more deviant, and thus even more peculiar.

The deviant galaxies also have a stronger FIR luminosity and a 
lower oxygen abundance.  
This is shown in Fig.~\ref{devtf}, which
give the difference between the observed \vm\ and that given by
Eq.~\ref{tfk} as a function of FIR luminosity (\lfir\ in \lsun)
and oxygen abundance (O/H) respectively.
Only the data with an accuracy in \lvm\ better than 0.1
and an inclination larger than 30$^{\circ}$ are
plotted.  The same holds true for the departure from the TF
relation in the $B$ band.  These trends are reinforced if one plots
data with an accuracy in \lvm\ better than 0.2 (instead of 0.1).
It thus appears that the most deviant galaxies are the youngest ones and 
those with the highest star formation rates, which is proportional to \lfir.

\section{Conclusions}
\label{conclu}

The first objective of this paper was to search for differences in HI 
content between Mrk IRAS and ``normal'' galaxies, and between barred 
and unbarred galaxies. 
We compared the HI properties of two homogeneous samples of Mrk IRAS 
(mainly starburst) and normal galaxies, with a further distinction 
between barred and unbarred objects. The main results of this
comparative study can be summarized as follows. 

Taking into account selection effects due to different distributions
in distances and morphological type between normal and Mrk IRAS
galaxies, we find that unbarred galaxies have more HI gas than barred ones.
This is probably the most important result of this paper. 
Our interpretation is that the bar contributes efficiently to transforming the
neutral hydrogen into molecular form, by funneling it into 
denser and dust-rich regions where most CO is detected : 
the leading edge of the bar (Sheth et al. 2002)
and subsequently the central regions of the galaxy,
where most star formation takes place.  Note, however, that
the bars of the Mrk IRAS galaxies are too young to have
contributed to the actual central starburst
(Coziol et al. 2000; Consid\`ere et al. 2000). 

We also find that the Mrk IRAS galaxies have more neutral 
gas than normal galaxies.  Although less robust, this result
suggests that these galaxies have transformed less HI into
stars because they are young, as testified by their metallicity
(e.g. Coziol et al. 1997).

The other objective of this study was to investigate how the 
Mrk IRAS galaxies behave with respect to
the TF relation, which is a fundamental scaling relation
between the galaxy mass and its large-scale dynamics. 
We found that a significant fraction of Mrk IRAS galaxies strongly
deviate from the TF relation defined for normal galaxies, both in the 
$B$ and the $K$ bands. This is the other striking result 
of this study and, as already pointed out by Coziol et al.\ (2000), 
an important property of starburst galaxies. However, the
interpretation of these deviations is not straightforward,
because we have no detailed kinematic information on these galaxies. 

Peculiar HI distribution and kinematics seem to be quite common among 
star-forming galaxies (e.g., Taramopoulos, Payne \& Briggs 2001; 
Meurer, Staveley-Smith \& Killeen 1998; Swaters, Sancisi \& van der 
Hulst 1997; Hurt, Turner \& Ho 1996; Kobulnicky \& Skillman 1995). 
The motion of HI gas in the vertical direction has
been studied in several nearby spiral galaxies. The general
interpretation of these vertical motions is that they are the result of the 
formation and expansion of supershells around starburst nuclei due to
stellar winds from massive OB stars and/or supernova explosions 
(e.g. Sancisi et al. 2001). Finally, Barton et al. (2001) have recently
shown that a significant fraction of star-forming galaxies in close
pairs deviate from the TF relation, with lower values of
\lvm. Although most of these outliers show signs of recent star
formation, Barton et al. (2001) suggest that gas dynamical effects are
probably the dominant cause of their anomalous TF properties, some 
of them showing signs of recent gas infall after a close galaxy-galaxy 
pass.

Most of the deviant galaxies in Table~\ref{devTF} are isolated, suggesting
that strong gravitational interactions are not at the origin of the peculiar
HI kinematics, except perhaps for Mrk 201 and 313.  We assume that the 
HI gas has been recently accreted by
these galaxies in minor merger events, and has not settled to dynamical
equilibrium (lopsided profiles), or is partly in non-planar
motion induced by superwinds (gaussian/centrally peaked profiles). 
A single minor merger is not sufficient to raise
the HI mass excess by 0.15, since it would require that the typical merging
dwarf contribute 40\% of the HI mass of the galaxy.  We thus have to
assume either that our galaxies frequently suffered minor mergers
(say 4 to 7 in their lifetime), which
seems unprobable in view of their generally low metallicity
(and thus age), or that they
were initially gas-rich; a combination of the two possibilities is
probably the correct scenario, since, as pointed out by the referee,
gas-rich disks are likely to respond to mergers by more vigorous starburst. 

How can we interpret these results in terms of the evolutionary stage of
starburst galaxies? 

In the framework of the hierarchical formation of
galaxies, gravitational interactions play a fundamental role in
galaxy evolution, primarily through minor galaxy-galaxy encounters. 
In the local universe, starburst nucleus galaxies are certainly the best
examples of such processes. They frequently show signatures of recent
interactions (e.g., Contini 1996) and have systematically lower metallicities
than normal galaxies of similar masses and morphological types (Coziol 
et al. 1997), suggesting a particularly early evolutionary
stage. Moreover, recent studies have shown that their bar structures
were formed recently (age $< 1$ Gyr; Consid\`ere et al.\ 2000) mainly
because these galaxies are still forming their disks (Coziol et al.\
2000). A detailed and exhaustive study of their chemical abundances
further revealed that the star formation history of starburst nucleus
galaxies is probably dominated by a succession of starbursts extending
over a few Gyr (Coziol 1996; Coziol et al.\ 1999; Mouhcine \& Contini
2002). 
The most natural explanation for such a star formation scenario
involves the trigger of starbursts by gravitational interactions, mainly 
through minor mergers. 

The results of this paper, which concerns both the global gas content and the
large-scale HI dynamics in starburst galaxies, further support this 
evolutionary scenario. Starburst galaxies contain more neutral gas
than normal ones because they recently underwent several minor mergers
which increased their reservoir of HI gas and triggered nuclear starbursts.
A significant fraction of these galaxies show
strong deviations from the TF relations because the recently accreted
HI gas has not had time to become stable in a regularly rotating disk and 
reach dynamical equilibrium. The fact that the most deviant
galaxies are those with the highest star formation rates and the lowest 
metallicities is not surprizing. The level of dynamical perturbations
in the HI kinematics (measured by the departure from the TF relation), 
as well as the level of induced star formation (measured by \lfir) 
should scale in principle with the intensity of the latest 
gravitational interaction undergone by the galaxy. Following the
well-known scaling relation between metallicity and mass, 
metal-poor galaxies have small masses and are thus more affected by 
interactions than massive ones, explaining the observed trend in
Fig.~\ref{devtf}. 

As predicted by numerical simulations, low-level gravitational 
interactions are the best way to
initiate instabilities in the stellar galactic disk and thus to form
a bar. This could explain why bars are so ``young'' in starburst
galaxies: their disks are still forming and they suffered recent
minor interactions. The difference in HI content between
barred and unbarred galaxies -- the latter being richer -- could be
explained by the efficiency of bars in driving gas toward denser and 
dust-richer regions of galaxies, where it is transformed into molecular clouds 
on its way to forming stars. Following this scenario, barred galaxies should 
contain more molecular gas than unbarred ones. We are investigating this
prediction thanks to new CO observations of Mrk starburst
galaxies (Contini et al., in preparation).

Over the past few years, considerable progress have been achieved in
our understanding of galaxy formation and evolution. In particular,
deep photometric and spectroscopic surveys performed on the largest 
telescopes revelead populations of luminous star-forming galaxies up
to redshifts 3 -- 4. Even if it remains unclear whether these objects 
are representative of the whole galaxy population that exists in the
distant universe, the starburst phenomenon appears now as a key
process in galaxy evolution.

\begin{acknowledgements}
We thank the technical staff of Observatoire de Nan\c cay for their help in
acquiring the data, and Gilles Theureau for help with the absolute
calibration.  We acknowledge with thanks detailed constructive comments
from an anonymous referee, which considerably improved the paper.
This research has made use of LEDA (the Lyon-Meudon
Extragalactic Database; {\tt http://leda.univ-lyon1.fr}), and of 
NED (the NASA/IPAC Extragalactic Database; 
{\tt http://nedwww.ipac.caltech.edu}) which is operated by the Jet Propulsion
Laboratory, California Institute of Technology, under contract with the
National Aeronautics and Space Administration.
\end{acknowledgements}

\begin{table*}
\caption[]{HI data for Markarian galaxies observed at the Nan\c cay 
Radiotelescope.  The Mrk number is in col.\ 1,
the radial heliocentric HI velocity and its error (in \kms) in cols.\ 2 and 3,
the rms noise in the profile (in Jy) in col.\ 4, the signal-to-noise
ratio (SNR) in col.\ 5,
the measured linewidths at 20 and 50\% of maximum intensity
(in \kms) in cols.\ 6 and 7. When no line was detected, the
listed $W50$ was estimated from the Tully-Fisher relation. 
The HI flux \fhi\ (or an upper limit) and its uncertainty (in \jykms) 
are given in cols.\ 8 and 9.
The last galaxy of that Table is UGCA 304, the companion of Mrk 1337.}
{\scriptsize 
\begin{flushleft}
\begin{tabular}{rrrrrrrrrrr}
\noalign{\smallskip}
\hline
\hline
\noalign{\smallskip}
Mrk & t & incl & $V_{\odot}$ & $\Delta V_{\odot}$ & rms & SNR 
& $W20$  & $W50$
& $F_{\rm HI}$ & $\Delta F_{\rm HI}$\\
\noalign{\smallskip}
& & [deg]& [\kms] & [\kms] & [mJy] &   & [\kms] & [\kms] 
& [\jykms] & [\jykms] \\
\noalign{\smallskip}
(1)&(2)&(3)&(4)&(5)&(6)&(7)&(8)&(9)&(10)&(11)\\
\noalign{\smallskip}
\hline
  21& 3.6& 58&   8433& 18& 7.1& 3.4&301&236&   1.37& 0.80\\ 
  38& 3.4& 70&  10813& 28& 6.7& 3.2&287&148&   1.70& 0.82\\ 
  42& 2.7& 38& [7200]&   &10.1&    &   &[132]&$<$3.98&   \\ 
  54& 5.1& 69&  13450& 16& 4.8& 3.2&225&185&   1.64& 0.68\\ 
  78&    & 50&[11194]&   & 9.1&    &   &[236]&$<$6.45&   \\ 
  84& 3.6& 62&   6098& 19& 5.7& 5.0&310&158&   4.30& 0.99\\ 
  90& 3.8& 29&   4253& 13& 4.5& 4.0&264&221&   2.74& 0.78\\ 
 103& 4.8& 29& [9353]&   & 8.8&    &   &[167]&$<$4.40&   \\ 
 109& 8.0& 50&   9054&  7& 5.6& 8.0&308&251&   7.47& 1.01\\ 
 122& 2.0& 63& [6538]&   &10.1&    &   &[289]&$<$8.78&   \\ 
 141&-4.9&   &[12265]&   &12.1&    &   &[141]&$<$5.11&   \\ 
 144& 4.8& 25& [8253]&   & 9.6&    &   &[118]&$<$3.39&   \\ 
 152& 2.7& 68& [6896]&   & 7.6&    &   &[277]&$<$6.28&   \\ 
 236& 2.7& 47&  14895& 19&10.4& 4.4&259&142&   3.77& 1.32\\ 
 237& 3.4& 38&   8937& 13& 3.9& 3.3&186&160&   2.13& 0.70\\ 
 238& 4.9& 27&  14884& 31& 6.3& 3.3&283&110&   1.90& 0.84\\ 
 262&    & 71&   8918& 25& 8.1& 2.8&208&129&   2.44& 1.11\\ 
 264&    &   &[18746]&   &10.6&    &   &[154]&$<$4.87&   \\ 
 271& 3.1& 42&   7541&   & 6.2& 3.4&340&243&   2.41& 0.74\\ 
 288& 4.9& 74& [7500]&   & 9.2&    &   &[227]&$<$6.25&   \\ 
 291& 1.1& 47&[10552]&   & 6.1&    &   &[198]&$<$3.61&   \\ 
 306& 4.0& 57&   5609&  8& 5.4& 8.9&257&176&   7.89& 0.96\\ 
 311& 4.8& 19&   9190& 15& 4.4& 3.7&310&261&   1.37& 0.55\\ 
 333&-1.8& 37&   4811&  7& 5.1& 2.6&117&111&   1.15& 0.66\\ 
 359& 2.7& 38& [5043]&   & 4.5&    &   &[162]&$<$2.19&   \\ 
 373&    & 53&   5903& 26& 5.1& 3.1&229&121&   1.91& 0.78\\ 
 374& 2.7& 69&  12780&  8&15.5& 6.9&121& 74&   8.54& 1.94\\ 
 382& 4.6& 25&  10099& 16& 5.1& 3.2&189&149&   1.15& 0.59\\ 
 412& 7.5& 71&   4479&  6& 4.9& 7.4&166&127&   4.09& 0.73\\ 
 414& 1.7& 57&  11350&  7& 4.0& 4.7&461&443&   2.36& 0.62\\ 
 445& 4.2& 61& [4600]&   & 6.4&    &   &[182]&$<$3.49&   \\ 
 455& 9.7& 61&  10175& 37& 4.7& 2.9&375&180&   1.75& 0.74\\ 
 474&    & 38&[10732]&   & 2.6&    &   &[222]&$<$1.71&   \\ 
 551& 3.6& 44&[15061]&   &17.8&    &   &[228]&$<$12.20&   \\ 
 552&-0.3& 32&   4349&  8&17.2& 7.3&299&237&  23.81& 3.32\\ 
 558&-1.3& 43& [3957]&   &14.4&    &   &[160]&$<$6.92&   \\ 
 592& 3.1& 52&   7716&  5& 9.0& 5.1&181&167&   5.73& 1.39\\ 
 596& 4.2& 32&  11594& 20&12.9& 5.6&432&222&  12.19& 2.35\\ 
 607& 1.1& 81&   2663& 16& 3.8& 4.4&133& 51&   0.74& 0.35\\ 
 618& 3.0& 44&  10684& 20& 4.0& 4.3&308&192&   2.84& 0.71\\ 
 623& 3.8& 49&  12617& 10& 3.6& 5.7&128& 74&   1.04& 0.37\\ 
 626& 0.3&  5&   3933&  7& 5.7& 4.7& 66& 47&   1.02& 0.49\\ 
 661& 4.8& 58&[10606]&   &11.1&    &   &[261]&$<$8.70&   \\ 
 665& 5.5& 57&   7630& 17& 6.7& 3.3&228&174&   2.18& 0.92\\ 
 688& 2.0& 49&  11539& 10& 4.2& 6.7&260&190&   4.63& 0.74\\ 
 712& 3.6& 63&   4565& 12& 7.7& 7.4&323&202&   9.25& 1.38\\ 
 718& 5.4& 26&   8517& 14& 4.6& 4.3&241&182&   2.09& 0.67\\ 
 721& 4.9& 38&   9621& 16& 3.2& 5.7&243& 96&   1.26& 0.37\\ 
 731&-1.3& 49&   1453& 10& 3.9& 4.1&212&183&   2.32& 0.63\\ 
 766& 1.0& 39& [3819]&   & 4.6&    &   & [94]&$<$1.28&   \\ 
 785& 1.3& 42&  14750& 23& 4.8& 2.9&307&235&   1.42& 0.68\\ 
 804& 4.9& 33&   5365& 16& 3.7& 4.3&253&179&   1.89& 0.58\\ 
 827& 4.1& 74&   5464&  8& 7.3& 7.7&346&282&  11.64& 1.47\\ 
 834& 2.0& 27&  10802& 20& 6.4& 5.2&377&204&   3.88& 0.96\\ 
 851& 1.7& 18&  10665& 13& 3.8& 4.1&197&153&   0.90& 0.39\\ 
 860& 3.4& 62&   6868& 10& 4.1& 4.6&140&104&   1.50& 0.52\\ 
 861& 4.9& 21&   4398& 14& 4.7& 5.0&217&146&   2.64& 0.69\\ 
 874& 3.4& 48&   4080&  7& 4.7& 7.2&168&127&   4.01& 0.72\\ 
 881& 5.5& 49&   8682& 13& 6.4& 4.7&226&164&   3.40& 0.96\\ 
 898& 3.1& 54&   5072& 13& 4.2& 4.0&296&256&   1.94& 0.63\\ 
 899& 3.8& 27&   5525& 13& 8.2& 4.8&170&108&   4.28& 1.19\\ 
 905&-1.8& 53&   4703&  5&12.9& 7.8&135&110&   9.87& 1.78\\ 
 910& 1.1& 38&   4910& 16& 5.6& 3.5&282&232&   1.95& 0.78\\ 
 912& 0.1& 63& [4845]&   &15.6&    &   &[238]&$<$11.13&   \\ 
 922& 4.9&   &   6574& 10& 8.2& 5.4&145& 95&   3.18& 0.99\\ 
 928&-0.9& 40&   7250& 13& 4.7& 5.9&145& 57&   1.73& 0.51\\ 
 929& 3.4& 49&   5259& 16& 5.3& 4.3&296&223&   4.43& 1.04\\ 
 936& 3.1& 73& [9059]&   &14.8&    &   &[343]&$<$15.21&   \\ 
 937& 5.3& 28&   8846& 16& 5.6& 5.7&205& 68&   2.39& 0.68\\ 
 943& 3.9& 61&   5186& 19& 3.9& 4.5&245&128&   1.80& 0.56\\ 
\noalign{\smallskip}
\noalign{\smallskip}
\hline
\hline
\end{tabular}
\end{flushleft}

}
\label{Tdata}
\end{table*} 

\begin{table*}
\caption[]{HI data for Markarian galaxies observed at the Nan\c cay 
Radiotelescope ({\it continued})}
{\scriptsize 
\begin{flushleft}
\begin{tabular}{rrrrrrrrrrr}
\noalign{\smallskip}
\hline
\hline
\noalign{\smallskip}
Mrk & t & incl & $V_{\odot}$ & $\Delta V_{\odot}$ & rms & SNR 
& $W20$ & $W50$
& $F_{\rm HI}$ & $\Delta F_{\rm HI}$\\                                  
\noalign{\smallskip}
& & [deg] & [\kms] & [\kms] & [mJy] &   
& [\kms] & [\kms] 
& [\jykms] & [\jykms] \\
\noalign{\smallskip}
(1)&(2)&(3)&(4)&(5)&(6)&(7)&(8)&(9)&(10)&(11)\\
\noalign{\smallskip}
\hline
\noalign{\smallskip}
 955& 4.2& 61&  10490& 26& 6.2& 3.0&246&149&   1.47& 0.78\\ 
 966& 0.4& 47&   5335& 16& 5.5& 2.3&156&133&   0.75& 0.60\\ 
 968& 4.0& 24&   4553&  5&10.7& 8.0&170&137&  11.01& 1.67\\ 
 997& 4.2& 63&   5795&  5& 6.3&11.7&320&243&  14.53& 1.23\\ 
1026& 2.0& 45&   3994&  8& 8.3& 8.2&271&205&  12.87& 1.59\\ 
1039& 5.0& 75&   2111&  5& 6.7& 8.3&161&134&   6.16& 0.97\\ 
1043& 0.2& 48&   2435& 13& 4.0& 4.0&172&133&   1.86& 0.60\\ 
1058& 4.8& 66& [5138]&   & 4.8&    &   &[215]&$<$3.07&   \\ 
1064&-1.9& 39&   4644&  5& 6.9&12.7&173&101&   9.34& 1.00\\ 
1066&-1.2& 68&   3605& 30& 3.9& 3.5&280&104&   1.11& 0.53\\ 
1076& 4.0& 67& [7232]&   & 4.1&    &   &[229]&$<$2.84&   \\ 
1081& 1.7& 20&   5327& 10& 5.0& 6.4&284&222&   5.76& 0.94\\ 
1085& 4.2& 59&   5127& 16& 5.6& 6.0&354&217&   5.22& 0.97\\ 
1086& 2.7& 42&   9043& 13& 5.2& 5.6&287&208&   4.30& 0.89\\ 
1092& 3.1& 76&   3356&  3& 5.9&14.7&338&311&  19.90& 1.22\\ 
1104& 2.1& 27&   1991& 37&11.0& 1.4&155&109&  16.85& 4.99\\ 
1116& 4.9& 37&   7823& 16&11.9& 3.0&283&246&   2.68& 1.43\\ 
1149& 5.0& 63&   6292&  5& 4.4& 5.8&276&262&   4.29& 0.79\\ 
1180&    & 46& [4534]&   & 6.0&    &   &[213]&$<$3.82&   \\ 
1197& 4.2& 57&   4958& 18& 3.4& 3.5&298&235&   2.11& 0.64\\ 
1198& 3.5& 69&   7635& 10& 8.4& 4.9&509&466&   9.63& 1.79\\ 
1199& 2.7& 43&   4085& 17& 6.2& 4.9&171& 60&   1.78& 0.67\\ 
1200& 2.7& 26&   7788&  8& 4.3& 6.2&293&253&   4.45& 0.77\\ 
1270& 5.0& 44&   8307& 14& 7.9& 5.8&299&196&   7.60& 1.41\\ 
1273&    & 55&   8184& 50& 6.1& 2.2&382&179&   1.29& 0.71\\ 
1302& 3.0& 62&   5588&  2& 5.5&14.8&108& 92&   7.00& 0.71\\ 
1337& 2.1& 54&   2556&  5& 5.8& 8.0&251&231&   7.91& 1.05\\ 
1344&-2.0& 31&   3170&  7& 7.2& 9.5&182&101&   6.44& 0.97\\ 
1356& 1.7& 45&  11872& 25& 6.4& 2.8&157& 83&   0.80& 0.60\\ 
1361&    & 25& [6768]&   & 6.2&    &   &[185]&$<$3.44&   \\ 
1363& 1.1& 34&   2743&  3& 6.8& 9.9&206&197&   9.46& 1.09\\ 
1403& 4.8& 53&   8693& 11& 5.4& 5.1&191&144&   2.54& 0.73\\ 
1404& 4.8& 27&[10687]&   & 9.4&    &   &[168]&$<$4.75&   \\ 
1425& 4.5& 61&   7480&  7& 5.0& 5.1&280&258&   5.25& 1.00\\ 
1452& 4.9& 41&   6680& 12& 5.6& 5.5&200&131&   4.18& 0.92\\ 
UA 304&  & 90&   2176&  8& 5.7&12.7&245& 66&   3.94& 0.59\\ 
\noalign{\smallskip}
\hline
\hline
\end{tabular}
\end{flushleft}

}
\label{Tdatb}
\end{table*} 

\begin{table*}
\caption[]{Integrated quantities for the Mrk IRAS
  galaxies. The Mrk number is in col.\ 1, the morphological type 
(SA = unbarred, SB = barred) in col.\ 2, the spectral type (stb =
starburst; sy = Seyfert; st? and sy? are classifications from an IRAS 
color index as explained in text) in col.\ 3, the numerical
morphological type $t$ (RC3) in col.\ 4, 
the inclination (in degrees) in col.\ 5,
the isophotal diameter $d_{\rm c}$ and 
its uncertainty (in log of 0.1\arcmin) in cols.\ 6 and 7, the maximum
rotation velocity \vm\ and its uncertainty in log(\kms) in cols.\ 8 and
9, the distance $D$ (in Mpc) in col.\ 10, the $K$-band magnitude in
col.\ 11, the oxygen abundance (O/H) with respect to solar in col.\ 12, 
the total HI mass and its uncertainty (in \msun) in cols.\ 13 and 14, 
the total FIR luminosity (in \lsun) in col.\ 15 and the absolute blue magnitude
\mabs\ in col.\ 16.}
{\scriptsize 
\begin{flushleft}
\begin{tabular}{rrlrrrrrrrrrrrrr}
\noalign{\smallskip}
\hline
\hline
\noalign{\smallskip}
Mrk & Bar & Spe &$t$&incl.&\multicolumn{2}{c}{$\log(d_{\rm c})$}
& \multicolumn{2}{c}{$\log(V_{\rm m})$} & $D$ & $m_{\rm K}$ &
O/H & \multicolumn{2}{c}{$\log(M_{\rm HI})$} &$\log(L_{\rm FIR})$& $M_{\rm B}$ \\
\noalign{\smallskip}
&&&& [deg]&\multicolumn{2}{c}{[0.1\arcmin]}& \multicolumn{2}{c}{[\kms]}
&[Mpc]& [mag]
& [$\odot$] & \multicolumn{2}{c}{[\msun]} & [\lsun] & [mag] \\
\noalign{\smallskip}
(1)&(2)&(3)&(4)&(5)&(6)&(7)&(8)&(9)&(10)&(11)&(12)&(13)&(14)&(15)&(16)\\
\noalign{\smallskip}
\hline
\noalign{\smallskip}
   1&SB&sy2& 3.1&   56.8&  0.871&  0.051&  2.122&  0.029&  65.9& 11.17&  0.11&  9.57&  0.24& 10.12&   -19.73\\
   2&SB&stb& 0.6&   27.2&  0.850&  0.051&  2.251&  0.232&  75.7& 10.64&  1.61&  9.47&  0.30& 10.64&   -20.75\\
   4&SB&stb& 6.0&   63.2&  1.283&  0.042&  2.111&  0.045&  73.9& 10.41&      &  9.58&  0.30& 10.30&   -20.81\\
   8&SA&stb& 4.1&   62.7&  0.943&  0.044&  2.067&  0.044&  51.3& 11.88&  0.22&  9.73&  0.27&  9.96&   -19.94\\
  10&SB&sy1& 3.1&   68.4&  1.229&  0.050&  2.430&  0.018& 119.6& 10.35&  0.65& 10.63&  0.10& 10.34&   -22.69\\
  12&SA&stb& 5.0&   38.0&  1.056&  0.052&  2.245&  0.088&  56.1& 10.21&  0.80& 10.08&  0.10& 10.22&   -21.10\\
  13&SB&stb& 3.1&   31.5&  1.023&  0.084&  2.544&  0.162&  21.9& 11.80&  0.61&  9.55&  0.10&  8.59&   -17.76\\
  21&SB&stb& 3.6&   58.2&  0.983&  0.054&  2.196&  0.096& 115.4& 11.11&  1.03&  9.63&  0.58& 10.07&   -21.04\\
  31&SB&stb& 2.5&   35.9&  0.936&  0.066&  2.146&  0.136& 107.1& 11.25&  1.70&  9.76&  0.10& 10.01&   -20.23\\
  38&SB&stb& 3.4&   70.2&  0.856&  0.056&  2.058&  0.204& 147.2& 12.25&  0.59&  9.94&  0.48&      &   -20.93\\
  52&SB&stb&-0.7&   63.5&  1.327&  0.046&  1.734&  0.058&  29.0&  9.78&  0.78&  9.00&  0.24&  9.72&   -19.37\\
  54&SA&stb& 5.1&   68.9&  0.881&  0.063&  2.029&  0.125& 181.3& 12.84&  0.25& 10.10&  0.41& 10.68&   -21.83\\
  58&SB&stb& 1.3&   31.0&  0.761&  0.048&  2.471&  0.112&  74.0& 12.28&  0.97&  8.69&  0.30&  9.48&   -19.36\\
  79&SB&sy2& 3.0&   28.2&  1.142&  0.052&  2.256&  0.208&  91.0&  9.80&      &  9.90&  0.08& 10.24&   -21.48\\
  84&SA&st?& 3.6&   62.2&  1.001&  0.073&  2.119&  0.120&  84.1& 11.14&      &  9.86&  0.23& 10.18&   -21.25\\
  86&SB&stb& 8.4&   35.0&  1.512&  0.293&  1.874&  0.280&   6.9&  9.46&  0.36&  8.65&  0.12&  8.66&   -17.36\\
  87&SB&stb& 1.0&   47.5&  1.118&  0.084&  2.154&  0.049&  51.1&  9.78&  0.38&  9.87&  0.30&  9.79&   -20.12\\
  90&SB&stb& 3.8&   29.4&  0.889&  0.075&  2.390&  0.044&  59.3& 10.64&      &  9.36&  0.28&  9.74&   -20.32\\
 101&SA&stb& 4.6&   17.6&  0.880&  0.097&  2.174&  0.545&  66.0& 10.94&  0.51&  9.69&  0.30&  9.87&   -20.42\\
 109&SB&stb& 8.0&   49.7&  0.529&  0.041&  2.256&  0.032& 123.7& 12.97&  0.51& 10.43&  0.14& 10.16&   -19.36\\
 114&SB&stb& 3.1&   28.7&  1.182&  0.046&  2.326&  0.167& 103.4& 10.80&  0.43& 10.30&  0.10& 10.42&   -20.61\\
 133&SB&stb& 4.0&   12.3&  1.054&  0.045&  2.499&  1.256&  31.2&  9.94&  1.74&  9.22&  0.12&  9.66&   -19.30\\
 158&SA&stb& 1.0&   67.7&  1.232&  0.045&  2.021&  0.018&  31.7&  9.59&      &  9.02&  0.08& 10.07&   -19.77\\
 161&SB&stb& 4.1&   52.6&  0.883&  0.071&  1.861&  0.103&  82.5& 10.75&  0.74&  9.99&  0.30& 10.39&   -21.14\\
 179&SB&stb& 5.3&   35.9&  1.163&  0.072&  2.110&  0.081&  47.7& 10.09&  1.44&  9.36&  0.14&  9.59&   -19.98\\
 185&SB&stb& 5.9&   42.6&  1.353&  0.042&  2.270&  0.061&  44.2&  9.70&      &  9.95&  0.13&  9.91&   -20.80\\
 188&SB&stb& 5.3&   39.7&  1.264&  0.056&  2.280&  0.095&  35.3&  9.25&  2.76&  9.70&  0.10& 10.01&   -20.43\\
 201&SA&stb& 9.8&   60.1&  1.261&  0.071&  1.766&  0.093&  36.7&  9.63&  0.44&  9.33&  0.30& 10.59&   -20.48\\
 207&SA&st?& 1.0&   42.6&  1.116&  0.053&  1.908&  0.158&  36.8& 10.38&      &  9.19&  0.27&  9.67&   -19.70\\
 213&SB&stb& 1.1&   49.2&  1.210&  0.052&  2.294&  0.059&  45.0&  9.75&  0.37&  9.40&  0.33& 10.07&   -20.55\\
 236&SA&sy1& 2.7&   47.0&  0.819&  0.055&  2.130&  0.117& 210.0& 11.84&      & 10.58&  0.35& 10.20&   -21.10\\
 237&SA&stb& 3.4&   38.4&  0.732&  0.082&  2.129&  0.080& 123.8& 12.01&  0.61&  9.89&  0.33& 10.53&   -19.84\\
 238&SA&stb& 4.9&   26.6&  0.933&  0.046&  2.315&  0.125& 201.7& 11.52&  0.84& 10.26&  0.44& 10.92&   -21.79\\
 256&SA&st?& 5.0&   42.1&  1.050&  0.066&  2.164&  0.080&  45.6& 11.01&      &  9.64&  0.18&  9.97&   -20.07\\
 262&SB&stb&    &   71.3&  0.620&  0.050&  1.937&  0.241& 123.8& 13.15&  0.41&  9.95&  0.45&      &   -18.88\\
 271&SB&stb& 3.1&   41.5&  0.906&  0.048&  2.339&  0.088& 104.3& 10.29&  0.47&  9.79&  0.31& 10.49&   -21.21\\
 281&SB&stb& 3.1&   40.0&  1.422&  0.057&  2.366&  0.081&  32.9&  8.72&      &  9.75&  0.35& 10.00&   -20.64\\
 286&SA&stb& 4.2&   30.3&  0.951&  0.044&  2.355&  0.156& 105.1& 10.58&  0.44& 10.29&  0.10& 10.84&   -21.39\\
 297&SA&stb& 5.0&   41.9&  0.932&  0.059&  2.481&  0.084&  65.2&      &  0.56&  9.93&  0.29& 10.62&   -21.34\\
 300&SB&stb&-0.1&   59.5&  0.847&  0.102&  2.006&  0.042& 159.9& 11.36&  0.74&  9.86&  0.30& 10.45&   -21.27\\
 306&SB&stb& 4.0&   56.6&  1.045&  0.043&  2.106&  0.052&  76.2& 12.06&  0.63& 10.03&  0.12& 10.04&   -20.24\\
 307&SB&stb& 4.6&   35.0&  1.068&  0.047&  2.004&  0.088&  75.6& 10.46&  1.00&  9.82&  0.18& 10.26&   -21.19\\
 308&SA&stb& 0.0&   52.7&  0.858&  0.090&  2.345&  0.047&  96.5& 11.58&  0.19&  9.94&  0.28& 10.46&   -20.39\\
 311&SB&st?& 4.8&   19.0&  0.641&  0.114&  2.636&  0.029& 123.5& 11.25&      &  9.69&  0.40& 10.65&   -20.91\\
 313&SB&sy2&-1.9&   61.9&  1.092&  0.074&  1.914&  0.073&  27.3&  9.54&      &  9.55&  0.38&  9.55&   -19.18\\
 318&SA&sy?& 3.8&   37.1&  0.910&  0.038&  2.218&  0.097&  60.0& 10.92&      &  9.74&  0.25&  9.94&   -20.14\\
 319&SB&stb& 1.1&   53.4&  1.099&  0.076&  2.053&  0.057& 109.3& 10.63&  0.44&  9.98&  0.22& 10.88&   -21.67\\
 321&SA&sy?& 5.9&   15.3&  1.046&  0.063&  2.632&  0.467& 129.3& 10.32&      & 10.51&  0.21& 10.88&   -22.27\\
 323&SB&stb& 5.9&   47.0&  1.097&  0.079&  2.275&  0.042&  58.6&  9.88&      &  9.48&  0.20& 10.29&   -21.01\\
 325&SA&st?& 4.9&   43.4&  1.142&  0.096&  1.972&  0.209&  46.8& 10.73&  0.38&  9.65&  0.23& 10.19&   -20.33\\
 326&SB&stb& 4.0&   55.7&  1.191&  0.075&  2.166&  0.044&  48.8& 10.26&  1.83&  9.76&  0.24& 10.12&   -20.04\\
 331&SA&st?& 1.0&   48.6&  0.957&  0.063&  2.099&  0.050&  73.7&  9.88&      &  9.98&  0.21& 11.08&   -20.21\\
 332&SB&stb& 4.2&   24.3&  1.165&  0.050&  1.935&  0.223&  33.2&  9.37&  0.81&  9.05&  0.13&  9.94&   -20.14\\
 339&SB&stb&    &   21.5&  0.962&  0.085&  2.286&  0.039&  71.0& 11.74&      &  9.73&  0.30&  9.68&   -20.13\\
 353&SB&stb& 2.2&   65.0&  0.950&  0.046&  1.927&  0.030&  63.2& 10.31&  0.55&  9.32&  0.28& 10.35&   -19.88\\
 358&SB&sy1& 4.0&   42.2&  0.896&  0.081&  2.396&  0.051& 182.1& 11.27&      & 10.16&  0.24& 10.37&   -21.51\\
 359&SB&sy1& 2.7&   38.0&  0.889&  0.068&  1.933&  0.114&  68.3& 10.46&  0.30&  8.85&  0.30&  9.91&   -20.43\\
 363&SA&stb&-2.0&   36.5&  0.793&  0.115&  2.063&  0.074&  40.1& 10.93&  0.41&  9.39&  0.32&  9.75&   -19.23\\
 373&SB&stb&    &   53.1&  0.890&  0.050&  2.026&  0.204&  81.3& 11.29&  0.64&  9.47&  0.41& 10.23&   -19.77\\
 374&SB&sy1& 2.7&   69.1&  0.824&  0.127&  1.673&  0.142& 176.8& 11.11&      & 10.80&  0.23&      &   -22.38\\
 382&SB&sy1& 4.6&   25.3&  0.836&  0.078&  2.285&  0.069& 137.0& 11.47&      &  9.71&  0.51&  9.91&   -20.42\\
 384&SB&stb& 3.0&   54.8&  1.151&  0.042&  2.283&  0.032&  63.4&  9.82&      &  9.29&  0.19& 10.38&   -20.82\\
 386&SB&stb& 3.8&   67.2&  1.275&  0.047&  2.337&  0.027&  48.5&  9.39&      &  9.90&  0.19&  9.51&   -20.84\\
 389&SA&st?& 1.0&   29.2&  1.276&  0.070&  2.408&  0.158&  64.1&  9.31&      & 10.01&  0.16&  9.78&   -21.12\\
 391&SA&stb& 1.1&   53.3&  1.091&  0.065&  2.230&  0.060&  55.0& 10.27&      &  9.39&  0.25&  9.86&   -20.25\\
 401&SB&stb& 0.5&   34.9&  1.031&  0.045&  1.978&  0.156&  24.1& 10.31&  1.47&  8.83&  0.27&  9.32&   -18.15\\
 412&SB&stb& 7.5&   70.5&  0.764&  0.074&  1.869&  0.068&  61.7& 12.67&  0.26&  9.56&  0.18&  9.52&   -19.22\\
 414&SA&sy?& 1.7&   57.1&  0.805&  0.060&  2.424&  0.022& 153.9& 11.50&      & 10.12&  0.26& 10.39&   -21.30\\
 430&SA&stb& 0.0&   59.4&  1.310&  0.054&  2.195&  0.027&  81.1&  9.81&      &  9.96&  0.10&  9.94&   -21.73\\
 432&SA&stb& 9.9&   90.0&  1.223&  0.096&  2.146&  0.030&  47.0& 10.74&  0.56& 10.30&  0.15& 10.09&   -20.26\\
 446&SB&stb& 3.1&   40.8&  1.148&  0.083&  1.948&  0.132&  97.0& 10.35&  0.80&  9.92&  0.17& 10.31&   -21.40\\
\hline
\end{tabular}
\end{flushleft}

}
\label{Tdera}
\end{table*} 

\begin{table*}
\caption[]{Integrated quantities for the Mrk IRAS
  galaxies ({\it continued})}
{\scriptsize 
\begin{flushleft}
\begin{tabular}{rrlrrrrrrrrrrrrr}
\noalign{\smallskip}
\hline
\hline
\noalign{\smallskip}
Mrk & Bar & Spe &$t$&incl.&\multicolumn{2}{c}{$\log(d_{\rm c})$}
& \multicolumn{2}{c}{$\log(V_{\rm m})$} & $D$ & $m_{\rm K}$ &
O/H & \multicolumn{2}{c}{$\log(M_{\rm HI})$} &$\log(L_{\rm FIR})$& $M_{\rm B}$ \\
\noalign{\smallskip}
&&&& [deg]&\multicolumn{2}{c}{[0.1\arcmin]}& \multicolumn{2}{c}{[\kms]}
&[Mpc]& [mag]
& [$\odot$] & \multicolumn{2}{c}{[\msun]} & [\lsun] & [mag] \\
\noalign{\smallskip}
(1)&(2)&(3)&(4)&(5)&(6)&(7)&(8)&(9)&(10)&(11)&(12)&(13)&(14)&(15)&(16)\\
\noalign{\smallskip}
\hline
\noalign{\smallskip}
 455&SA&st?& 9.7&   61.4&  0.733&  0.138&  2.201&  0.194& 138.9& 12.36&      &  9.90&  0.42& 10.66&   -20.70\\
 471&SB&sy2& 1.2&   53.1&  0.946&  0.056&  2.124&  0.061& 139.6& 10.66&      &  9.79&  0.19& 10.40&   -21.69\\
 479&SA&stb& 3.9&   60.8&  1.013&  0.062&  2.219&  0.039&  83.0& 10.98&  0.50& 10.05&  0.19& 10.27&   -21.17\\
 489&SB&stb& 2.3&   51.3&  0.915&  0.092&  1.991&  0.058& 130.9& 11.30&  0.57& 10.58&  0.10& 10.74&   -21.55\\
 493&SB&sy1& 3.1&   42.1&  1.000&  0.073&  1.220&  0.166& 128.9& 10.98&      &  9.81&  0.21& 10.23&   -20.93\\
 496&SA&stb& 1.7&   68.7&  0.796&  0.223&  2.150&  0.022& 121.8& 11.02&  0.45& 10.07&  0.30& 11.14&   -21.66\\
 527&SB&stb& 1.1&   56.9&  1.153&  0.083&  1.541&  0.105&  47.6& 10.07&      &  9.22&  0.12& 10.17&   -19.90\\
 529&SA&stb&-1.0&   90.0&  1.129&  0.051&  2.024&  0.035&  47.1& 11.09&  0.25& 10.18&  0.26&  9.72&   -18.95\\
 531&SA&st?&-1.8&   58.3&  1.211&  0.060&  2.163&  0.031&  48.0&  9.70&      &  8.37&  0.30& 10.19&   -20.03\\
 533&SB&sy2& 3.9&   24.1&  0.997&  0.141&  2.621&  0.202& 117.3&  9.79&      & 10.35&  0.20& 11.03&   -21.92\\
 534&SA&stb&-1.3&   59.2&  1.180&  0.145&  2.164&  0.081&  68.7& 10.02&  0.54&  9.84&  0.45& 10.67&   -21.11\\
 538&SB&stb& 3.1&   50.0&  1.299&  0.085&  2.070&  0.086&  37.3&  9.76&  0.34&  9.80&  0.17& 10.29&   -20.42\\
 545&SB&stb& 1.2&   42.4&  1.279&  0.080&  2.446&  0.079&  62.2&  8.88&  0.92&  9.89&  0.18& 10.73&   -21.47\\
 552&SB&st?&-0.3&   32.4&  0.618&  0.147&  2.399&  0.027&  58.0& 11.09&      & 10.28&  0.14& 10.25&   -19.87\\
 554&SA&sy?&-2.0&   41.6&  1.169&  0.098&  2.105&  0.074&  70.4& 11.06&      &  9.73&  0.30&  9.70&   -20.09\\
 555&SA&stb& 3.2&   29.2&  1.106&  0.057&  2.141&  0.237&  53.7&  9.66&      &  9.66&  0.30& 10.29&   -21.22\\
 565&SA&sy?&-2.0&   43.2&  1.316&  0.053&  2.352&  0.079&  72.5&  9.75&      &  9.22&  0.30& 10.19&   -21.54\\
 571&SB&stb& 3.4&   50.4&  1.216&  0.081&  2.342&  0.047&  68.1& 10.41&  1.80&  9.80&  0.30& 10.03&   -20.48\\
 572&SA&st?& 7.9&   83.0&  1.079&  0.057&  2.299&  0.034&  65.4& 10.47&      &  9.09&  0.34& 10.17&   -20.21\\
 575&SB&stb& 1.0&   32.4&  0.991&  0.051&  2.077&  0.121&  73.3& 10.25&  0.82&  9.69&  0.21& 10.39&   -20.48\\
 592&SB&stb& 3.1&   52.0&  0.994&  0.142&  2.022&  0.040& 102.3& 11.52&      & 10.15&  0.24& 10.36&   -20.47\\
 593&SB&stb& 5.3&   24.6&  1.113&  0.042&  2.363&  0.230& 110.1& 10.15&      & 10.22&  0.17& 10.30&   -21.50\\
 596&SA&sy2& 4.2&   31.8&  0.932&  0.115&  2.495&  0.053& 142.6& 10.89&      & 10.74&  0.19& 10.44&   -21.70\\
 602&SB&stb& 4.0&   34.0&  1.102&  0.080&  2.294&  0.157&  37.2& 10.11&  1.34&  9.70&  0.24&  9.86&   -19.52\\
 607&SA&sy2& 1.1&   80.9&  1.275&  0.076&  1.582&  0.349&  34.8&  9.36&      &  8.33&  0.47&  9.57&   -19.80\\
 617&SB&stb& 4.9&   35.0&  1.174&  0.073&       &       &  62.0&  9.55&  0.69&  9.53&  0.30& 11.20&   -21.34\\
 618&SB&sy1& 3.0&   43.9&  0.964&  0.076&  2.255&  0.093& 137.8& 10.45&  0.81& 10.10&  0.25& 10.87&   -22.44\\
 620&SB&sy2& 1.0&   52.4&  1.488&  0.074&  2.323&  0.057&  27.3&  8.48&  0.16&  9.30&  0.07&  9.84&   -20.16\\
 623&SA&sy?& 3.8&   49.2&  0.924&  0.051&  1.781&  0.138& 169.4& 12.53&      &  9.85&  0.36& 10.63&   -21.24\\
 626&SA&sy?& 0.3&    5.0&  0.826&  0.062&  2.390&  0.024&  53.8& 10.65&      &  8.84&  0.48&  9.67&   -19.64\\
 665&SB&stb& 5.5&   57.2&  0.998&  0.118&  2.069&  0.121& 108.3& 11.26&  0.41&  9.78&  0.42& 10.16&   -21.13\\
 685&SA&st?& 2.8&   63.3&  0.816&  0.043&  1.783&  0.071&  62.1& 12.73&  0.22&  9.66&  0.34&  9.65&   -18.96\\
 686&SB&sy2& 1.7&   58.9&  1.084&  0.110&  2.292&  0.046&  59.3&  9.67&  0.12&  9.17&  0.30&  9.63&   -20.34\\
 688&SA&st?& 2.0&   49.5&  0.733&  0.085&  2.168&  0.057& 144.7& 12.04&  0.64& 10.36&  0.16& 10.30&   -20.69\\
 691&SB&stb& 5.1&   59.4&  1.208&  0.043&  1.954&  0.044&  46.2& 10.42&  1.50&  9.91&  0.22& 10.12&   -20.72\\
 701&SA&   & 1.1&   39.3&  1.063&  0.051&  2.134&  0.141&  70.7& 11.24&      &  9.57&  0.10&  9.99&   -19.75\\
 703&SB&stb& 2.1&   40.6&  1.310&  0.061&  1.890&  0.158&  51.0&  9.51&      & 10.01&  0.11& 10.19&   -20.57\\
 708&SB&stb& 4.1&   74.4&  1.341&  0.039&  2.052&  0.015&  27.1&  9.68&  1.48&  9.15&  0.30&  9.76&   -18.95\\
 710&SB&stb& 2.5&   55.4&  1.341&  0.032&  2.030&  0.034&  20.2&  9.96&  1.24&  9.13&  0.12&  9.20&   -18.21\\
 712&SB&stb& 3.6&   62.9&  1.036&  0.070&  2.169&  0.068&  61.9& 11.82&  0.35&  9.92&  0.15&  9.71&   -20.70\\
 718&SA&stb& 5.4&   26.2&  0.795&  0.051&  2.374&  0.049& 112.1& 10.58&  2.33&  9.79&  0.32& 10.46&   -20.67\\
 721&SA&stb& 4.9&   38.0&  0.720&  0.053&  2.112&  0.103& 128.6& 11.97&  1.40&  9.69&  0.29& 10.24&   -20.39\\
 731&SB&stb&-1.3&   49.3&  1.317&  0.073&  2.102&  0.066&  19.4&  9.70&  1.94&  8.31&  0.27&  9.21&   -17.79\\
 739&SA&sy1& 3.8&   33.7&  0.789&  0.080&  2.429&  0.187& 120.3& 10.45&  0.57&  9.41&  0.30& 10.49&   -21.17\\
 752&SB&stb& 4.1&   37.1&  0.971&  0.056&  2.017&  0.083&  82.0& 11.70&  0.44&  9.86&  0.23&  9.90&   -20.38\\
 759&SB&stb& 5.2&   40.0&  1.271&  0.128&  2.173&  0.057&  30.2&  9.63&      &  9.70&  0.20&  9.81&   -19.94\\
 769&SA&stb& 1.0&   65.7&  1.302&  0.077&  1.999&  0.025&  24.2&  9.49&  0.30&  9.72&  0.33&  9.85&   -19.80\\
 781&SB&stb& 4.6&   41.6&  1.305&  0.059&  2.161&  0.078&  38.9&  9.87&  0.49&  9.51&  0.20&  9.68&   -20.15\\
 785&SA&   & 1.3&   42.2&  0.628&  0.094&  2.304&  0.095& 198.8& 11.64&      & 10.12&  0.48& 10.72&   -22.25\\
 799&SB&stb& 3.1&   51.6&  1.337&  0.043&  2.290&  0.044&  43.2&  8.98&  2.27&  9.59&  0.18& 10.48&   -20.86\\
 804&SB&stb& 4.9&   33.4&  0.813&  0.073&  2.291&  0.068&  73.1& 13.04&  0.31&  9.38&  0.31&  9.76&   -19.97\\
 809&SA&stb& 5.3&   59.5&  0.982&  0.126&  1.795&  0.037& 104.0&      &      &  9.96&  0.30& 10.48&   -20.57\\
 814&SB&stb& 3.2&   56.9&  1.018&  0.288&  2.279&  0.076&  54.7& 10.38&  0.38&  9.76&  0.19&  9.84&   -18.75\\
 827&SA&st?& 4.1&   74.0&  1.042&  0.040&  2.210&  0.041&  75.9& 11.21&      & 10.20&  0.13& 10.01&   -20.57\\
 834&SA&st?& 2.0&   27.3&  1.021&  0.086&  2.508&  0.052& 147.4& 10.50&      & 10.30&  0.25& 10.94&   -21.81\\
 839&SA&st?& 1.7&   48.2&  1.232&  0.116&  1.851&  0.104&  55.6& 10.50&      &  9.79&  0.40& 10.36&   -21.12\\
 851&SA&sy?& 1.7&   18.1&  0.969&  0.050&  2.438&  0.040& 144.7& 11.82&      &  9.65&  0.43& 10.46&   -20.90\\
 853&SA&st?& 4.9&    0.0&  0.954&  0.107&       &       &  84.4& 11.62&      &  9.76&  0.30& 10.08&   -20.62\\
 860&SB&stb& 3.4&   61.8&  0.645&  0.165&  1.811&  0.129&  94.3&      &  0.25&  9.50&  0.35& 10.43&   -20.28\\
 861&SB&stb& 4.9&   20.8&  0.756&  0.118&  2.393&  0.047&  60.9& 10.63&      &  9.36&  0.26&  9.85&   -19.91\\
 871&SB&sy1& 5.2&   60.3&  0.816&  0.072&  2.424&  0.025& 137.1& 10.90&      &  9.88&  0.30& 10.31&   -21.72\\
 874&SB&stb& 3.4&   48.4&  0.745&  0.057&  1.977&  0.062&  59.0& 13.20&  0.34&  9.52&  0.18&  9.33&   -18.69\\
 881&SA&st?& 5.5&   48.6&  0.624&  0.094&  2.102&  0.086& 118.9& 11.80&  1.75& 10.05&  0.28& 10.52&   -20.71\\
 898&SB&stb& 3.1&   54.1&  0.987&  0.059&  2.224&  0.065&  69.1& 10.89&  1.78&  9.34&  0.32&  9.78&   -19.66\\
 899&SA&sy?& 3.8&   27.2&  0.977&  0.046&  2.164&  0.074&  72.7& 11.28&      &  9.73&  0.28&  9.82&   -20.47\\
 905&SA&st?&-1.8&   53.5&  1.071&  0.056&  1.855&  0.058&  65.3& 10.07&      & 10.00&  0.18&  9.95&   -20.06\\
 906&SA&st?&-0.1&   41.3&  1.261&  0.066&  2.264&  0.071&  86.5& 10.07&      & 10.03&  0.30& 10.32&   -20.58\\
 907&SA&st?& 7.7&   78.8&  1.289&  0.125&  1.891&  0.032&  18.1& 10.25&  0.54&  8.81&  0.16&  9.16&   -19.54\\
 908&SA&sy?& 2.7&   61.6&  1.177&  0.067&  1.978&  0.034&  19.0& 10.83&      &  8.87&  0.18&  8.72&   -18.94\\
 910&SA&st?& 1.1&   38.4&  0.916&  0.140&  2.313&  0.065&  65.2& 10.39&      &  9.29&  0.40&  9.90&   -20.38\\
 917&SB&sy2& 1.0&   21.5&  0.998&  0.085&  2.458&  0.410&  99.7& 10.54&      &  9.38&  0.30& 10.74&   -21.05\\
 922&SA&st?& 4.9&    0.0&  0.659&  0.165&       &       &  78.4& 11.32&      &  9.66&  0.31&  9.98&   -20.30\\
 928&SA&sy2&-0.9&   40.0&  1.133&  0.122&  1.823&  0.163&  97.2& 10.75&      &  9.59&  0.29& 11.01&   -20.93\\
 929&SA&st?& 3.4&   49.5&  0.766&  0.095&  2.232&  0.078&  72.6& 11.26&      &  9.74&  0.23& 10.04&   -20.75\\
 937&SA&sy2& 5.3&   27.5&  0.826&  0.048&  2.105&  0.105& 118.0& 11.19&      &  9.89&  0.28& 10.15&   -21.06\\
 943&SA&sy?& 3.9&   60.8&  0.693&  0.128&  2.017&  0.152&  69.1& 11.78&      &  9.31&  0.31&  9.82&   -19.67\\
\hline
\end{tabular}
\end{flushleft}

}
\label{Tderb}
\end{table*} 

\begin{table*}
\caption[]{Integrated quantities for the Mrk IRAS
  galaxies ({\it continued})}
{\scriptsize 
\begin{flushleft}
\begin{tabular}{rrlrrrrrrrrrrrrr}
\noalign{\smallskip}
\hline
\hline
\noalign{\smallskip}
Mrk & Bar & Spe &$t$&incl.&\multicolumn{2}{c}{$\log(d_{\rm c})$}
& \multicolumn{2}{c}{$\log(V_{\rm m})$} & $D$ & $m_{\rm K}$ &
O/H & \multicolumn{2}{c}{$\log(M_{\rm HI})$} &$\log(L_{\rm FIR})$& $M_{\rm B}$ \\
\noalign{\smallskip}
&&&& [deg]&\multicolumn{2}{c}{[0.1\arcmin]}& \multicolumn{2}{c}{[\kms]}
&[Mpc]& [mag]
& [$\odot$] & \multicolumn{2}{c}{[\msun]} & [\lsun] & [mag] \\
\noalign{\smallskip}
(1)&(2)&(3)&(4)&(5)&(6)&(7)&(8)&(9)&(10)&(11)&(12)&(13)&(14)&(15)&(16)\\
\noalign{\smallskip}
\hline
\noalign{\smallskip}
 955&SB&sy2& 4.2&   60.6&  0.935&  0.064&  2.046&  0.195& 139.0& 10.76&      &  9.83&  0.53& 10.49&   -21.21\\
 958&SA&   & 4.0&   86.1&  1.079&  0.046&  2.140&  0.012&  84.8& 11.43&      &  9.84&  0.30&  9.91&   -20.64\\
 960&SA&stb&-1.9&   71.9&  0.900&  0.050&  1.981&  0.060&  84.1& 11.93&  0.30&  9.75&  0.10& 10.14&   -20.20\\
 966&SA&st?& 0.4&   46.9&  0.952&  0.065&  1.971&  0.143&  70.4& 10.83&      &  8.94&  0.80& 10.09&   -19.62\\
 968&SB&stb& 4.0&   24.1&  1.094&  0.048&  2.256&  0.023&  58.8& 10.17&      &  9.96&  0.15&  9.65&   -20.47\\
 984&SB&lin& 7.9&   69.8&  1.049&  0.072&  2.284&  0.037& 190.4& 10.59&      & 10.03&  0.21& 10.58&   -22.04\\
 987&SA&st?& 4.6&   65.9&  1.131&  0.077&  2.281&  0.024&  66.5& 10.39&      &  9.08&  0.30& 10.15&   -20.80\\
 997&SA&sy?& 4.2&   63.3&  1.024&  0.040&  2.197&  0.026&  76.3& 11.22&      & 10.30&  0.08& 10.09&   -20.41\\
1002&SA&stb&-1.5&   38.6&  1.078&  0.117&  2.183&  0.088&  41.9& 10.10&  0.73&  9.21&  0.23& 10.07&   -19.47\\
1003&SA&sy?& 2.7&   53.1&  0.911&  0.074&  1.946&  0.046&  48.4& 10.90&      &  8.97&  0.30&  9.36&   -19.56\\
1009&SB&stb& 3.1&   63.0&  1.297&  0.188&  2.264&  0.038&  56.8& 10.25&  0.78&  9.86&  0.19&  9.82&   -20.85\\
1026&SB&stb& 2.0&   44.8&  1.152&  0.049&  2.221&  0.040&  51.5& 10.46&  0.65&  9.91&  0.12&  9.79&   -20.29\\
1027&SA&stb& 8.8&   45.6&  0.844&  0.099&  2.457&  0.056& 120.6&      &  0.59& 10.03&  0.26& 11.05&   -20.73\\
1039&SA&stb& 5.0&   75.2&  1.181&  0.069&  1.860&  0.057&  26.4& 11.84&  0.18&  9.01&  0.16&  9.12&   -18.64\\
1043&SA&stb& 0.3&   48.3&  1.143&  0.053&  1.992&  0.110&  31.7& 10.22&  0.91&  8.65&  0.32&  9.39&   -19.10\\
1050&SB&stb& 1.1&   67.5&  1.085&  0.073&  2.099&  0.047&  67.0& 10.12&      &  9.59&  0.11& 10.52&   -20.54\\
1063&SA&stb& 9.1&   70.6&  1.265&  0.046&  1.977&  0.032&  18.5& 10.51&  0.27&  9.53&  0.29&  9.18&   -19.09\\
1064&SA&st?&-1.9&   38.8&  1.060&  0.049&  2.014&  0.040&  59.5& 10.83&      &  9.89&  0.11& 10.15&   -20.07\\
1066&SB&sy2&-1.2&   68.2&  1.214&  0.107&  1.991&  0.255&  49.7&  9.79&  0.25&  8.81&  0.48& 10.53&   -19.78\\
1067&SB&stb& 1.7&   40.2&  1.152&  0.058&  2.356&  0.064&  57.5&  9.87&      &  9.66&  0.10& 10.07&   -19.98\\
1068&SA&   & 9.1&   90.0&  1.123&  0.043&  2.035&  0.027&  68.3& 11.13&      & 10.26&  0.30&  9.96&   -20.21\\
1073&SB&sy2& 3.1&   32.8&  1.047&  0.049&  2.474&  0.134&  94.9& 10.11&  0.62&  9.60&  0.30& 11.01&   -21.16\\
1080&SB&   & 8.2&   90.0&  1.560&  0.092&  1.879&  0.031&  10.2& 10.37&      &  9.32&  0.06&  8.43&   -19.03\\
1081&SA&sy?& 1.7&   20.4&  0.976&  0.047&  2.563&  0.023&  70.3& 10.64&      &  9.83&  0.16&  9.89&   -20.35\\
1085&SA&st?& 4.2&   59.2&  1.058&  0.045&  2.224&  0.080&  65.2& 10.45&      &  9.72&  0.19&  9.89&   -20.89\\
1086&SB&stb& 2.7&   42.1&  0.953&  0.045&  2.264&  0.059& 113.2& 11.58&  0.79& 10.11&  0.21& 10.21&   -20.49\\
1087&SA&stb&-1.8&   47.8&  1.009&  0.066&  2.193&  0.067& 110.2& 11.86&  0.34& 10.22&  0.18& 10.77&   -21.13\\
1088&SB&stb& 0.1&   20.7&  1.244&  0.041&  2.654&  0.361&  60.4&  9.20&  1.38&  9.43&  0.22& 10.53&   -21.63\\
1089&SA&stb& 8.9&   64.7&  0.930&  0.219&  1.911&  0.089&  52.7& 11.84&  0.39& 10.18&  0.12& 10.19&   -18.92\\
1092&SB&st?& 3.1&   75.6&  1.244&  0.078&  2.215&  0.015&  43.0& 10.31&      &  9.94&  0.06&  9.32&   -21.20\\
1093&SA&stb& 1.1&   47.5&  1.076&  0.123&  2.362&  0.082&  57.9& 10.19&  2.05&  9.93&  0.30& 10.63&   -20.06\\
1104&SA&st?& 2.1&   26.5&  0.892&  0.082&  2.139&  0.224&  29.8& 11.54&      &  9.55&  0.30&  9.23&   -17.56\\
1116&SA&   & 4.9&   36.8&  0.968&  0.103&  2.333&  0.062& 109.2& 10.56&      &  9.88&  0.53& 10.96&   -21.36\\
1118&SA&st?& 3.0&   48.6&  1.285&  0.040&  2.360&  0.124&  44.1&  9.65&      & 10.13&  0.12&  9.75&   -20.11\\
1124&SA&sy?& 4.8&   39.0&  1.134&  0.038&  2.091&  0.058&  66.4& 11.10&      & 10.00&  0.12&  9.58&   -20.32\\
1127&SA&stb& 3.0&   75.8&  1.196&  0.041&  2.324&  0.012& 101.9& 10.49&  0.06& 10.13&  0.30& 10.24&   -21.19\\
1137&SA&st?& 4.3&   28.8&  0.866&  0.076&  2.292&  0.160& 103.5& 11.86&      &  9.89&  0.30& 10.26&   -20.30\\
1149&SB&stb& 5.0&   63.0&  1.021&  0.060&  2.166&  0.028&  82.9& 11.61&  1.07&  9.84&  0.18&  9.88&   -20.53\\
1157&SB&sy2&-0.1&   38.5&  1.165&  0.064&  2.086&  0.095&  62.3& 10.06&      &  9.37&  0.28& 10.05&   -20.33\\
1171&SA&stb& 5.3&   49.1&  1.214&  0.057&  1.667&  0.115&  70.5&  9.68&  2.40&  9.98&  0.11& 10.29&   -21.12\\
1194&SB&stb&-2.0&   53.8&  1.246&  0.044&  2.251&  0.044&  58.8&  9.23&      &  9.46&  0.19& 10.55&   -20.75\\
1197&SA&sy?& 4.2&   57.5&  0.799&  0.049&  2.197&  0.095&  66.7& 10.84&      &  9.34&  0.30&  9.91&   -19.63\\
1198&SA&sy?& 3.5&   68.8&  1.065&  0.051&  2.414&  0.032& 102.8& 10.26&      & 10.38&  0.19& 10.28&   -21.07\\
1199&SA&stb& 2.7&   42.9&  1.023&  0.076&  1.859&  0.196&  55.0& 10.16&  1.14&  9.10&  0.38& 10.45&   -20.79\\
1200&SB&stb& 2.7&   26.4&  0.962&  0.107&  2.486&  0.022& 104.7& 10.35&      & 10.06&  0.17& 10.39&   -20.85\\
1225&SA&sy?& 2.3&   66.5&  1.042&  0.170&  2.264&  0.020&  50.9& 10.92&      &  9.56&  0.30&  9.14&   -19.31\\
1233&SA&st?& 3.6&   57.5&  0.839&  0.053&  2.172&  0.058&  64.6& 11.18&      &  9.77&  0.27& 10.24&   -19.18\\
1235&SA&st?& 5.0&   52.5&  1.122&  0.045&  1.989&  0.100&  65.1& 10.37&      &  9.67&  0.10&  9.90&   -20.94\\
1270&SA&sy?& 5.0&   43.8&  1.086&  0.050&  2.250&  0.066& 110.7& 11.70&      & 10.34&  0.19& 10.24&   -20.84\\
1273&SB&stb& 0.0&   54.9&  0.851&  0.061&  2.234&  0.243& 110.1& 11.37&      &  9.57&  0.55& 10.13&   -20.33\\
1291&SB&sy2& 3.9&   33.2&  1.436&  0.039&  2.367&  0.097&  48.5&  9.52&      & 10.09&  0.10&  9.89&   -21.14\\
1302&SB&stb& 3.0&   62.5&  1.003&  0.055&  1.710&  0.032&  74.7& 10.94&  0.90&  9.96&  0.10&  9.68&   -19.54\\
1304&SA&stb& 3.0&   57.6&  1.053&  0.195&  2.137&  0.083&  73.4& 11.34&  0.53&  9.75&  0.30& 10.43&   -20.16\\
1326&SB&stb& 5.9&   28.9&  1.217&  0.031&  2.114&  0.107&  19.4& 10.96&  1.68&  8.64&  0.20&  8.85&   -18.56\\
1330&SB&sy1& 3.0&   54.9&  1.475&  0.094&  2.313&  0.091&  33.3&  8.11&      &  9.34&  0.15&  9.71&   -20.31\\
1333&SA&sy2& 2.9&   78.3&  1.216&  0.051&  2.306&  0.015&  37.7&  9.46&      &  9.47&  0.10&  9.83&   -19.54\\
1337&SA&st?& 2.1&   54.3&  1.113&  0.056&  2.160&  0.029&  34.4& 10.50&      &  9.34&  0.13&  9.76&   -18.89\\
1341&SB&stb& 5.9&   50.6&  1.330&  0.037&  2.087&  0.052&  16.2& 10.29&  0.85&  9.01&  0.36&  9.07&   -18.69\\
1344&SB&stb&-2.0&   30.8&  0.978&  0.050&  2.116&  0.045&  42.2& 10.37&  0.28&  9.44&  0.15&  9.80&   -18.72\\
1346&SB&stb& 6.5&   75.8&  1.217&  0.030&  1.831&  0.024&  15.4& 11.92&  0.20&  9.00&  0.09&  8.51&   -17.74\\
1356&SA&   & 1.7&   45.3&  0.842&  0.056&  1.894&  0.266& 159.1& 11.11&      &  9.68&  0.75& 10.50&   -21.07\\
1363&SB&stb& 1.1&   34.3&  1.278&  0.049&  2.236&  0.015&  36.7&  9.70&      &  9.48&  0.12&  9.46&   -20.38\\
1365&SB&stb&-2.0&   44.8&  0.985&  0.107&  2.101&  0.109&  75.5& 10.32&  0.42&  9.27&  0.30& 10.50&   -19.71\\
1379&SB&stb& 2.2&   33.6&  1.152&  0.079&  1.821&  0.153&  35.5&  9.63&  1.45&  9.45&  0.10&  9.93&   -19.99\\
1381&SA&st?& 4.0&   75.1&  1.030&  0.047&  2.281&  0.012& 109.5& 11.62&      &  9.91&  0.21& 10.24&   -20.49\\
1403&SA&sy1& 4.8&   53.4&  0.740&  0.045&  2.006&  0.091& 112.8& 12.44&      &  9.88&  0.29& 10.12&   -19.86\\
1425&SA&st?& 4.5&   60.5&  0.964&  0.067&  2.176&  0.039& 101.6& 10.85&      & 10.11&  0.19& 10.24&   -20.77\\
1452&SB&stb& 4.9&   40.9&  0.796&  0.072&  2.087&  0.082&  92.1& 11.54&  1.07&  9.92&  0.22&  9.85&   -19.85\\
1461&SA&sy?& 3.5&   52.2&  0.751&  0.071&  1.985&  0.040&  86.8& 12.56&      &  9.38&  0.27&  9.90&   -19.96\\
1466&SB&stb& 5.0&   47.5&  1.613&  0.065&  2.127&  0.073&  18.2&  8.79&  1.48&  9.67&  0.13&  9.48&   -19.55\\
1485&SB&stb& 3.6&   45.0&  1.430&  0.144&  2.282&  0.078&  33.9&  8.62&  2.12&  9.90&  0.13&  9.76&   -20.74\\
\hline
\end{tabular}
\end{flushleft}

}
\label{Tderc}
\end{table*} 

\end{document}